\documentclass[12pt]{article}
\usepackage{setspace}

\usepackage{amsthm,amsmath, amssymb, amsfonts, amscd, xspace}
\usepackage{epsfig, psfrag}
\usepackage{float, fancybox, fullpage}
\usepackage{delarray}
\usepackage{url}
\usepackage{authblk}

\usepackage{anysize}
\usepackage{multirow}
\usepackage[top=1in, bottom=1in, left=1in, right=1in]{geometry}
\usepackage{rotating}
\usepackage{latexsym,ifthen}
\usepackage{subfigure}
\usepackage{anysize}
\usepackage[round]{natbib} 
\usepackage{colortbl}

\newtheorem{lemma}{Lemma}
\newtheorem{theorem}{Theorem}

\newcommand{\bc}{\boldsymbol{c}}

\newcommand{\bK}{\boldsymbol{K}}
\newcommand{\bX}{\boldsymbol{X}}
\newcommand{\bx}{\boldsymbol{x}}

\newcommand{\bz}{\boldsymbol{z}}

\newcommand{\bbeta}{\boldsymbol{\beta}}
\newcommand{\balpha}{\boldsymbol{\alpha}}
\newcommand{\bsigma}{\boldsymbol{\sigma}}
\newcommand{\blambda}{\boldsymbol{\lambda}}

\newcommand{\bzero}{\boldsymbol{0}}

\newcommand{\mH}{\mathcal{H}}
\newcommand{\mbbR}{\mathbb{R}}

\newcommand{\mF}{\mathcal{F}}
\newcommand{\mG}{\mathcal{G}}

\newcommand{\mN}{\mathcal{N}}

\newcommand{\pr}{\textrm{pr}}

\def\argmin{\mathop{\rm argmin}}

\def\arginf{\mathop{\rm arginf}}

\begin{document}

\title{Double Sparsity Kernel Learning with Automatic Variable Selection and Data Extraction}

\author[1]{\vspace{-2 ex} Jingxiang Chen}
\author[2]{Chong Zhang}
\author[1,3]{Michael R. Kosorok}
\author[1,3,4,*]{Yufeng Liu\vspace{-2 ex}}
\affil[1]{Department of Biostatistics \vspace{-2 ex}}
\affil[3]{Department of Statistics and Operations Research \vspace{-2 ex}}
\affil[4]{Department of Genetics \vspace{-2 ex}}
\affil[ ]{University of North Carolina at Chapel Hill \vspace{-2 ex}}
\affil[2]{Department of Statistics and Actuarial Science \vspace{-2 ex}}
\affil[ ]{University of Waterloo \vspace{-2 ex}}
\affil[*]{Email: yfliu@email.unc.edu \vspace{-5 ex}}

\date{}

\maketitle

\begin{abstract}
Learning with Reproducing Kernel Hilbert Spaces (RKHS) has been widely used in many scientific disciplines. Because a RKHS can be very flexible, it is common to impose a regularization term in the optimization to prevent overfitting. Standard RKHS learning employs the squared norm penalty of the learning function. Despite its success, many challenges remain. In particular, one cannot directly use the squared norm penalty for variable selection or data extraction. Therefore, when there exists noise predictors, or the underlying function has a sparse representation in the dual space, the performance of standard RKHS learning can be suboptimal. In the literature,work has been proposed on how to perform variable selection in RKHS learning, and a data sparsity constraint was considered for data extraction. However, how to learn in a RKHS with both variable selection and data extraction simultaneously remains unclear. In this paper, we propose a unified RKHS learning method, namely, DOuble Sparsity Kernel (DOSK) learning, to overcome this challenge. An efficient algorithm is provided to solve the corresponding optimization problem. We prove that under certain conditions, our new method can asymptotically achieve variable selection consistency. Simulated and real data results demonstrate that DOSK is highly competitive among existing approaches for RKHS learning.
\end{abstract}


\section{Introduction}

Recent advances in technology have enabled scientists to collect massive datasets with high dimensions. For example, in online movie evaluation systems, the data sets can contain rating information from millions of users on thousands of movies. Extracting knowledge from such large data sets poses unprecedented challenges to existing learning techniques. To overcome new difficulties in mining big data sets, in the last few decades, many methodologies have been proposed in the machine learning literature. In this paper, we focus on supervised learning with one response variable. In particular, the learning goal is often to train a function using a training data set, such that for new observations, one can use this function to predict the unobserved responses. See \citet{hastie_elements_2011} for a comprehensive review of supervised learning techniques.


For many applications in supervised learning, appropriate variable selection is very important to the prediction performance of the estimated function. In particular, for real data sets, many predictors do not contain useful information with respect to the response. Hence, these redundant predictors should be excluded when we make further prediction. For instance, in classification problems, \citet{fan_sure_2008} showed that prediction using all variables may behave similarly to random guessing, due to the noise accumulation. How to perform variable selection has drawn much attention in the literature. Traditional methods for variable selection include forward and backward selections, among others. Recently, model fitting using sparse regularization has become very popular in the learning framework. The corresponding optimization problems of these techniques are equivalent to minimizing objective functions in the {\em loss} + {\em penalty} form. The loss term measures the goodness of fit of the estimated function, and the penalty term aims to select important variables in the learning problem, which further controls the complexity of the function space to prevent overfitting.

For different learning tasks, one uses different loss functions. For example, in least squares regression, one uses the squared error loss, and in standard Support Vector Machines \citep[SVM,][]{Vapnik92}, we use the hinge loss. For the penalty term, the choice depends on the corresponding functional space. In particular, if the response depends on the predictors linearly, linear learning should be used. Otherwise, one can employ various nonlinear learning methods such as splines \citep{boor_practical_2001} in regression. In this paper, we focus on learning in Reproducing Kernel Hilbert Spaces \citep[RKHS,][]{Aronszajn1950, Kimeldorf1971}. This is a very general setting, and many nonlinear learning techniques can be regarded as special cases of RKHS learning. For example, it covers penalized linear regression, additive spline models with or without interactions, and the entire family of smoothing splines. RKHS learning has been extensively used in the literature, and has achieved great successes. See, for example, \citet{Schlkopf2002}, \citet{kernelbook}, and \citet{hastie_elements_2011}.

For linear learning, variable selection with sparse regularization has been extensively studied. See, for example, \citet{tibshirani_regression_1996}, \citet{fan_variable_2001}, \citet{zou_regularization_2005}, \citet{wu_adaptive_2009}, \citet{zhang_nearly_2010}, \citet{fan2010selective}, and the references therein. For RKHS learning, however, the problem of variable selection has received much less attention. In the literature, \citet{guyon_gene_2002} suggested an extension of variable selection from linear learning to kernel learning using the Recursive Feature Elimination (RFE) approach. \citet{lin_component_2006} developed the Component Selection and Smoothing (COSSO), and proposed to use the sum of component norms as the sparse penalty, instead of the squared norm penalty in standard RKHS learning. \citet{zhang_linear_2011} proposed a structure selection method that can automatically determine whether the signal for one predictor is linear or nonlinear. Recently, \citet{allen_automatic_2012} developed an interesting framework of variable selection in RKHS learning. In particular, \citet{allen_automatic_2012} imposed a weight on each predictor, and proposed to train the model with a sparse penalty on the weight vector. When a fitted weight is zero, the corresponding predictor is regarded as unimportant in the learning problem, and is removed from further analysis. \citet{allen_automatic_2012} provided the Kernel Iterative Feature Extraction (KNIFE) algorithm to solve the corresponding optimization.

Despite the current progress in variable selection for RKHS learning, many challenges remain. First, theoretical properties of sparse penalties in linear learning have been well studied in the literature. For example, \citet{fan_variable_2001} and \citet{zou_adaptive_2006} proved the oracle property of their proposed methods, and \citet{zhao2006model} showed selection consistency for LASSO problems. In contrast, theoretical properties of existing variable selection approaches for RKHS learning are much less developed. In particular, it is desirable to explore conditions under which one can have consistency for kernel variable selection. Moreover, \citet{allen_automatic_2012} proposed to use the standard squared norm penalty on the learning function to avoid overfitting, besides the sparse penalty on the variable weight vector. However, as \citet{zhang_quantile_2015} pointed out, this approach uses all observations to represent the fitted function. This can lead to suboptimal prediction performance as the underlying function can be well approximated by a data sparse representation in the dual space \citep[see][and Section~\ref{sec:dosk} for more details]{zhang_quantile_2015}. Therefore, it can be beneficial to have a regularization method that can automatically select data points for RKHS learning. To circumvent this difficulty, \citet{zhang_quantile_2015} proposed a data sparsity constraint for data extraction. However, \citet{zhang_quantile_2015} did not consider the problem of kernel variable selection, and the data sparsity method can have suboptimal performance when there are noise covariates. Therefore, it is desirable to design a new method that can perform variable selection and data extraction simultaneously.

In this paper, we propose a new DOuble Sparsity Kernel (DOSK) learning method to fill this gap. We provide an efficient algorithm to solve the corresponding optimization problem. Through numerical examples, we show that our DOSK method can often select useful predictors accurately, and the sparsely represented functions can have very good prediction performance. Moreover, under some conditions, we prove that our DOSK method can enjoy many desirable statistical properties, including variable selection consistency.

The rest of the paper is organized as follows. In Section~\ref{sec:method}, we briefly introduce standard kernel learning methods, and discuss variable selection and data extraction for learning in a RKHS. Then, we propose our DOSK method, and develop our algorithm for the corresponding optimization problem. We establish some theoretical properties of DOSK, such as selection consistency, in Section~\ref{sec:theory}. Simulated and real data examples are used to demonstrate the effectiveness of our new method in Section~\ref{sec:numerical}. We provide some discussions in Section~\ref{sec:discussion}. All technical proofs are collected in the appendix.

\section{Methodology}
\label{sec:method}

We first give a brief review of standard kernel learning in Section~\ref{sec:standard}. Then we propose our DOSK method in Section~\ref{sec:dosk}. We discuss how to solve the corresponding optimization problem in Section~\ref{sec:optimization}.

\subsection{Standard Learning in RKHS}
\label{sec:standard}

Suppose each observation in the training data set $(\bx_i,y_i); \ i=1,\ldots,n$ is obtained from a fixed but unknown distribution $P(\bX,Y)$, where $\bX \in \mbbR^p$ is a vector of predictors, and $Y$ is the response. The learning goal is to find $f(\cdot)$ based on the training data set, so that for a new observation with only $\bx$ available, the prediction of $Y$ based on $f(\bx)$ can be accurate. For example, in regression, one often uses $f(\bx)$ to estimate the response $Y$, and in binary margin-based classification where $Y \in \{+1, -1\}$, one can let $\textrm{sign}\{ f(\bx) \}$ be the predicted label for $\bx$. For many learning problems, the goodness of fit of $f$ can be measured by a loss function $L\{Y, f(\bX)\}$. For different learning tasks, one uses different loss functions. For instance, in standard regression problems where the goal is to estimate the conditional mean of $Y$ with given $\bx$, it is common to use the squared error loss $L\{Y, f(\bX)\} = \{Y-f(\bX)\}^2$. In classification problems, one can use the hinge loss $L\{Y, f(\bX)\} = \{1-Yf(\bX)\}_+$ for support vector machines \citep[SVM,][]{Vapnik92}, and the deviance loss $L\{Y, f(\bX)\} = \log[1+\exp\{-Yf(\bX)\}]$ for logistic regression \citep{Wahba00}.

The optimization problem of a learning technique typically involves minimizing an objective function in the form of {\em loss} + {\em penalty}. In particular, the objective function can be written as
\begin{align}
\min_{f \in \mF} \frac{1}{n} \sum_{i=1}^n L\{y_i, f(\bx_i)\} + \lambda J(f),
\label{eq:general}
\end{align}
where $\mF$ is the function space for learning. Here the penalty term $J(f)$ regularizes $f(\cdot)$ in order to prevent overfitting, and the tuning parameter $\lambda$ balances $L(\cdot,\cdot)$ and $J(f)$ with the aim to achieve a good prediction performance. The choice of the penalty term varies based on $\mF$. For example, in standard linear regression, one often assumes that the conditional mean of $Y$ is a linear function of $\bx$, and it is common to use $\mF = \{f: \ f(\bx) = \bx^T \bbeta+\beta_0;\ \bbeta \in \mbbR^p, \beta_0 \in \mbbR\}$. There are many popular choices for $J(f)$ in the linear learning literature. See, for example, \citet{tibshirani_regression_1996}, \citet{fan_variable_2001}, \citet{zou_regularization_2005}, \citet{zhang_nearly_2010}, among others. If a linear function cannot estimate the response well, one often considers a nonlinear function space $\mF$. In this paper, we focus on learning in RKHS. For more details about RKHS, we refer the readers to \cite{wahba_spline_1990}, \citet{kernelbook}, and the references therein.

For learning in a RKHS $\mH$, it is common to use the squared norm penalty $J(f) = \|f\|_{\mH}^2$, where $\|f\|_{\mH}$ is the norm of $f$ in $\mH$. In other words, (\ref{eq:general}) can be written as
\begin{align}
\min_{f \in \mH} \frac{1}{n} \sum_{i=1}^n L\{y_i, f(\bx_i)\} + \lambda \|f\|_{\mH}^2.
\label{eq:generalL2}
\end{align}
\citet{Kimeldorf1971} showed that under mild conditions on $L$, the estimated function $\hat{f}$ from (\ref{eq:generalL2}) has the form $\hat{f}(\bx) = \sum_{i=1}^n \hat{\alpha}_i K(\bx_i,\bx)$, where $K(\cdot,\cdot)$ is the kernel function associated with $\mH$, $\bx_i$'s are the observed predictor vectors in the training data set, and $\alpha_i$'s are the parameters to estimate. Moreover, define $\bK$ to be the gram matrix with the $(i,j)$th element $K(\bx_i,\bx_j); \ i,j=1,\ldots,n$, and $\balpha=(\alpha_1,\ldots,\alpha_n)^T$. One can verify that the penalty $\|f\|_{\mH}$ in (\ref{eq:generalL2}) can be written as $\hat{\balpha}^T \bK \hat{\balpha}$. Consequently, (\ref{eq:generalL2}) is equivalent to the following problem,
\begin{align*}
\min_{\balpha \in \mbbR^n} \frac{1}{n} \sum_{i=1}^n L\{y_i, f(\bx_i)\} + \lambda \balpha^T \bK \balpha.
\end{align*}

In practice, however, many commonly used kernel spaces, for example the well known Gaussian RKHS, do not include offsets or intercepts \citep{Minh2010}. This can lead to suboptimal results for some learning problems. For instance, in quantile regression, if one is interested in estimating the $100\tau \%$ quantile of the response with $\tau$ close to $0$ or $1$, a regression function without an intercept can have inferior performance. Therefore, in this paper, we consider learning in RKHS with intercepts. In particular, in (\ref{eq:general}), we assume that $f = \tilde{f} + b \in \mH \oplus \mbbR$, and let $J(f)$ be the squared norm of $\tilde{f}$, where $\tilde{f}$ is the projection of $f$ onto $\mH$. The Representer's Theorem \citep{Kimeldorf1971} shows that under mild conditions, $\hat{f}(\bx) = \sum_{i=1}^n \hat{\alpha}_i K(\bx_i,\bx) + \hat{b}$, where $b$ is the intercept term, and $J(\hat{f}) = \hat{\balpha}^T \bK \hat{\balpha}$. Hence, for standard RKHS learning, the optimization problem (\ref{eq:generalL2}) with an intercept in $f$ can be written as
\begin{align}
\min_{\balpha \in \mbbR^n, b \in \mbbR} \frac{1}{n} \sum_{i=1}^n  L\{y_i, \sum_{j=1}^n \alpha_j K(\bx_i,\bx_j) + b\} + \lambda \balpha^T \bK \balpha.
\label{eq:generalL2new}
\end{align}

\subsection{Double Sparsity Kernel Learning}
\label{sec:dosk}

Despite the success of standard kernel learning methods, many challenges remain. First, the standard squared norm penalty cannot perform automatic variable selection. When the underlying signal depends only on a small fraction of the predictors (note that the corresponding relationship can be nonlinear), learning with all predictors can lead to overfitting, and consequently unsatisfactory results. In the literature, \citet{zhang_linear_2011} and \citet{allen_automatic_2012}, among others, proposed different methods for variable section in RKHS learning. In particular, to perform variable selection in kernel learning, \citet{allen_automatic_2012} proposed the idea of variable weighted kernel learning as follows. For a weight vector $\mathbf{w} \in \mbbR^p$ and any $\bx_1,\bx_2 \in \mbbR^p$, we define the variable weighted kernel function $K_{\mathbf{w}}(\bx_1,\bx_2) = K(\mathbf{w} \odot \bx_1, \mathbf{w} \odot \bx_2)$, where $\mathbf{w} \odot \bx$ denotes the element-wise product of vectors. In other words, the $j$th element of $\mathbf{w}$, $w_j$, represents the weight of the $j$th predictor of $\bX$ in the kernel function. For any positive definite kernel function $K$, one can verify by Mercer's Theorem that the newly defined variable weighted kernel $K_{\mathbf{w}}(\cdot,\cdot)$ naturally introduces a RKHS over the domain of $\bX$. For identifiability, we impose the constraint that $w_j \in [0,1]$ for all $j$. In the variable weighted kernel function, if $w_j=0$, then the $j$th predictor of $\bX$ has no impact on $f$ or the prediction. Therefore, one can impose an $L_1$ type penalty on the vector $\mathbf{w}$ to achieve variable selection in RKHS learning. In particular, \citet{allen_automatic_2012} proposed KNIFE for learning in a RKHS with variable selection, with the following optimization
\begin{align}
\underset{\alpha,b,w}{\mbox{min}}\left[\frac{1}{n} \sum_{i=1}^{n}L\big\{y_{i},
\sum_{j=1}^{n}K_{\mathbf{w}}(\bx_{i},\bx_{j})\alpha_{j}+b\big\}
+ \lambda_{1}\|\mathbf{w}\|_{1}
+\lambda_{2}\mathbf{\mathbf{\boldsymbol{\alpha}}}^{T}K_{\mathbf{w}}\mathbf{\boldsymbol{\alpha}}\right],
\label{eq:allen}
\end{align}
where $\lambda_1$ and $\lambda_2$ are tuning parameters, and $\mathbf{w}\in[0,1]^{p}$.

To better illustrate the variable weighted kernel function, we consider several commonly used RKHSs as examples. Define $x_{ik}$ to be the $k$th element of $\bx_i$. The linear variable weighted kernel is $K_{\mathbf{w}}(\bx_{i},\bx_{j})=\sum_{k=1}^{p}w_{k}^{2}x_{ik}x_{jk}$,
the polynomial variable weighted kernel is $K_{\mathbf{w}}(\bx_{i},\bx_{j})=\{c+\sum_{k=1}^{p}w_{k}^{2}(x_{ik}x_{jk})\}^{d}$
with $c\in\mathbb{R}$ and $d\in\mathbb{\mathbb{N}}$, the Gaussian variable
weighted kernel is $K_{\mathbf{w}}(\bx_{i},\bx_{j})=\exp\{-\gamma\sum_{k=1}^{p}(w_{k}x_{ik}-w_{k}x_{jk})^{2}\}$
with $\gamma\in\mathbb{R}^{+}$, and the Laplacian variable weighted kernel
is $K_{\mathbf{w}}(\bx_{i},\bx_{j})=\exp(-\gamma\sum_{k=1}^{p}|w_{k}x_{ik}-w_{k}x_{jk}|)$ with $\gamma\in\mathbb{R}^{+}$.

Recently, \citet{zhang_quantile_2015} showed that in some cases, using the squared norm penalty $\|\cdot\|_{\mH}^2$ for learning in RKHS can lead to suboptimal results. In particular, in a given learning problem, let $f^*(\bx)$ be the minimizer of the conditional expected loss. In other words, $f^*(\bx) = E [L \{Y, f(\bX)\} \mid \bX=\bx]$ for any $\bx$ (e.g., $f^*(\bx)$ is the conditional mean of $Y(\bx)$ in standard regression). \citet{zhang_quantile_2015} observed that if $f^*(\bx)$ can be well approximated by a function with a sparse representation in the RKHS (in other words, $f^*(\cdot)$ can be well approximated by $\sum_{i=1}^n \alpha_i K(\bx_i,\cdot) + b$ for only some nonzero $\alpha_i$), learning with the squared norm penalty can have the potential danger of overfitting. To overcome this difficulty, one can apply an $L_1$ penalty on the vector $\balpha$ for data selection of the estimated function. For RKHS learning problems, \citet{zhang_quantile_2015} proposed the data sparsity constraint with the following optimization
\begin{align}
\underset{\alpha,b }{\mbox{min}}\left[\frac{1}{n} \sum_{i=1}^{n}L\big\{y_{i},\sum_{j=1}^{n}K (\bx_{i},\bx_{j})\alpha_{j}+b\big\}
+ \lambda \|\balpha\|_{1}\right],
\label{eq:zhang}
\end{align}
where $K(\cdot,\cdot)$ is the standard kernel function and $\|\balpha\|_1=\sum_{i=1}^n{|\alpha_i|}$. Using the quantile regression as an example, \citet{zhang_quantile_2015} showed that, in certain cases, learning with the data sparsity constraint in (\ref{eq:zhang}) can improve the prediction performance.

Although data extraction was used in \citet{zhang_quantile_2015}, their method does not consider variable selection. Hence, when there are noise predictors in $\bx$, the proposed approach can be suboptimal. To our knowledge, not much work has been done on simultaneous data extraction and variable selection in the literature. To fill this gap, we propose our DOuble Sparsity Kernel learning (DOSK) method as follows
\begin{equation}
\underset{\balpha,b,w}{\mbox{min}}\left[\frac{1}{n} \sum_{i=1}^n L\big\{y_{i}, \sum_{j=1}^{n}K_{\mathbf{w}}(\bx_{i},\bx_{j})\alpha_{j}+b\big\}
+\lambda_{1}\|\mathbf{\boldsymbol{\alpha}}\|_{1}+\lambda_{2}\|\mathbf{w}\|_{1}
+\lambda_{3}\mathbf{\mathbf{\boldsymbol{\alpha}}}^{T}K_{\mathbf{w}}\mathbf{\boldsymbol{\alpha}}\right],
\label{eq:main}
\end{equation}
with $\lambda_i \ge 0; \ i=1,2,3$, $K_{\mathbf{w}}(\bx_1,\bx_2) = K(\mathbf{w} \odot \bx_1, \mathbf{w} \odot \bx_2)$ as defined earlier with $\mathbf{w}\in[0,1]^{p}$.

The framework of our DOSK (\ref{eq:main}) is very general, in the sense that it includes many existing approaches as special cases. In particular, when $\lambda_1=\lambda_2 = 0$, (\ref{eq:main}) reduces to the standard squared norm penalized kernel learning (\ref{eq:generalL2new}). When $\lambda_1 = 0$, (\ref{eq:main}) reduces to the KNIFE approach (\ref{eq:allen}) proposed by \citet{allen_automatic_2012}. If $\lambda_2=\lambda_3 = 0$, (\ref{eq:main}) becomes the data sparsity learning (\ref{eq:zhang}) in \citet{zhang_quantile_2015}. Because DOSK is a general framework of RKHS learning, one can use various approaches to solve the optimization problem (\ref{eq:main}), based on the choice of the loss function $L(\cdot,\cdot)$, $\mathbf{w}$ and $\lambda_l; \ l=1,2,3$. For example, in linear kernel learning with $\lambda_2 \ne 0$, one can verify that (\ref{eq:main}) is a biconvex problem with respect to $(\balpha^T, b)^T$ and $\mathbf{w}$, and can be solved by the alternate convex search algorithm \citep{gorski2007biconvex}. For more general DOSK problems, we propose a unified algorithm to solve (\ref{eq:main}) in the Section 2.3.

Note that although we impose multiple penalties in (\ref{eq:main}), our DOSK method can circumvent the difficulty of over-penalization by choosing $(\lambda_1, \lambda_2,  \lambda_3)$ carefully. In particular, in Section~\ref{sec:theory}, we show that if the tuning parameters are chosen appropriately, our DOSK method can enjoy many desirable theoretical properties.

\subsection{Computational Algorithm for DOSK}
\label{sec:optimization}

The major difficulty of solving the optimization (\ref{eq:main}) is that even $L$ is convex, the composite loss function $L\big\{y, \sum_{j=1}^{n}K_{\mathbf{w}}(\bx,\bx_{j})\alpha_{j}+b\big\}$ may not be convex with respect to $(\mathbf{w}^T,\balpha^T,b)^T$. Consequently, many existing algorithms for convex optimizations \citep{boyd_convex_2004} cannot be used directly. On the other hand, one can verify that if the loss function $L$ is convex, the optimization (\ref{eq:main}) is convex respect to $(\balpha^T,b)^T$ for a fixed $\mathbf{w}$. Hence, a natural way to circumvent the difficulty of non-convex optimization is to update $\mathbf{w}$ and $(\balpha^T,b)^T$ recursively. This, however, cannot be done directly, as for a general kernel function $K(\cdot,\cdot)$, $L\big\{y, \sum_{j=1}^{n}K_{\mathbf{w}}(\bx,\bx_{j})\alpha_{j}+b\big\}$ is not biconvex with respect to $\mathbf{w}$ and $(\balpha^T,b)^T$. One way to tackle this problem is that for fixed $(\balpha^T,b)^T$, we can find a linear approximation of the variable weighted kernel function $K_{\mathbf{w}}$ in a small neighbourhood of $(\mathbf{w}^T,\balpha^T,b)^T$ \citep{allen_automatic_2012}. Thus, to update $\mathbf{w}$, one can employ the linear approximation of $K_{\mathbf{w}}$ to make the corresponding objective function convex. Note that in the literature, the idea of local linear approximation has been widely used to solve optimizations for many learning problems. See, for example, \citet{an1997solving}, \citet{zou2008one}, \citet{lee2012multiple}, among others.

To introduce our algorithm for DOSK, we need some further notation. Let the objective function in (\ref{eq:main}) be $\phi( \mathbf{\boldsymbol{\alpha}} ,b,\mathbf{w})$. Define an $n\times p$ matrix $A(\mathbf{w})$, whose $i$th row is $\sum_{j=1}^{n}\alpha_{j} \nabla_{\mathbf{w}} K_{\mathbf{w}}(\bx_{i},\bx_{j})^{T}$, and an $n\times n$ matrix $B(\mathbf{w})$ with the $(i,j)$th element $B(i,j)=K_{\mathbf{w}}(\bx_{i},\bx_{j})-\nabla K_{\mathbf{w}}(\bx_{i},\bx_{j})^{T}\mathbf{w}$. Here $\nabla_{\mathbf{w}} K_{\mathbf{w}}(\bx_{i},\bx_{j})$ is the gradient vector of $K_{\mathbf{w}}(\bx_{i},\bx_{j})$ with respect to $\mathbf{w}$. By Taylor's expansion, one can verify that for $ \mathbf{w}_1 $ and $ \mathbf{w}_2 $, we have
\begin{align}
K_{\mathbf{w}_1} \balpha = A(\mathbf{w}_2) \mathbf{w}_1 + B(\mathbf{w}_2) \balpha + o(\|\mathbf{w}_1 - \mathbf{w}_2\|_2).
\label{eq:linear}
\end{align}
Define $\bc_{\mathbf{w}_2}(\mathbf{w}_1) = A(\mathbf{w}_2) \mathbf{w}_1 + B(\mathbf{w}_2) \balpha$, which is a linear function of $\mathbf{w}_1$. When $ \mathbf{w}_1 $ and $ \mathbf{w}_2 $ are close, we can use $\bc$ as the local linear approximation of $K_{\mathbf{w}} \balpha$ in our DOSK optimization algorithm. In particular, we outline the general algorithm to solve (\ref{eq:main}) in Algorithm 1 below.

\begin{center}
\begin{table*}[htb]
\hfill{}
\begin{tabular}{ l }
\hline
{\bf Algorithm 1}:  \\
\hline
\vspace{-3mm} \\
1. Initialize $\mathbf{w}^{\left(0\right)}$, $\mathbf{\mathbf{\boldsymbol{\alpha}}}^{\left(0\right)}$
and $b^{(0)}$ with $w_j \in [0,1]$ for $1 \le j \le p$. \\
\vspace{-3mm} \\
2. The $\boldsymbol{\alpha}$ step: fix $\mathbf{w}^{ (t-1 )}$ and $b^{(t-1)}$, and find
$\balpha^{(t)} = \argmin_{\balpha} \phi( \balpha, b^{(t-1)} ,\mathbf{w}^{ (t-1 )})$. \\
\hspace{4mm} The optimization problem is convex, and independent of the $\lambda_2\|\mathbf{w}\|_1$ term in (\ref{eq:main}). \\
\vspace{-3mm} \\
3. The $b$ step: fix $\mathbf{w}^{ (t-1 )}$ and $\balpha^{(t)}$, and find \\ \hspace{4mm}  $b^{(t)} = \argmin_b \frac{1}{n} \sum_{i=1}^{n} L\big\{y_{i}, \sum_{j=1}^{n}K_{\mathbf{w}^{(t-1)}}(\bx_{i},\bx_{j})\alpha_{j}^{(t)}+b\big\}$. This is a convex \\ \hspace{4mm} optimization with one parameter, and can be solved by standard methods. \\
\vspace{-3mm} \\
4. The $\mathbf{w}$ step: fix $b^{ (t )}$ and $\balpha^{(t)}$, and define $\bc_{\mathbf{w}^{(t-1)}}(\mathbf{w}) = A(\mathbf{w}^{(t-1)}) \mathbf{w} + B(\mathbf{w}^{(t-1)}) \balpha^{(t)}$. \\
\hspace{4mm} Let $\{ \bc_{\mathbf{w}^{(t-1)}}(\mathbf{w}) \}_i$ be the $i$th element of $\bc_{\mathbf{w}^{(t-1)}}(\mathbf{w})$. Under the constraint $\mathbf{w}^{(t)} \in [0,1]^p$, \\
\hspace{4mm} find \\
\hspace{4mm} $\mathbf{w}^{(t)} = \argmin_{\mathbf{w}} \frac{1}{n} \sum_{i=1}^n L [y_i, \{ \bc_{\mathbf{w}^{(t-1)}}(\mathbf{w}) \}_i + b^{(t)}] + \lambda_2 \|\mathbf{w}\|_1 + \lambda_3 \mathbf{w}^T A(\mathbf{w}^{(t-1)}) \balpha^{(t)} $. \\
\hspace{4mm} This is a standard quadratic programming problem. \\
5. Repeat steps 2-4 until convergence. \\
\vspace{-3mm} \\
\hline
\end{tabular}
\hfill{}
\label{algorithm:1}
\end{table*}
\end{center}

In the $\boldsymbol{\alpha}$ and $b$ steps in Algorithm 1, the corresponding objective functions are convex, therefore after updating the parameters, the value of $\phi$ decreases. On the other hand, in the $\mathbf{w}$ step, we replace the original objective function $\phi$ by its local linear approximation, and solve a quadratic programming problem. Denote the solution to this quadratic programming problem by $\mathbf{w}^{(QP)}$. In Algorithm 1, the updated $\mathbf{w}^{(t)}=\mathbf{w}^{(QP)}$ can have some distance from $\mathbf{w}^{(t-1)}$, hence the original $\phi$ function is not guaranteed to decrease. One possible way to overcome this difficulty is that in the $\mathbf{w}$ step, instead of having $\mathbf{w}^{(t)}=\mathbf{w}^{(QP)}$, we can treat $\mathbf{w}^{(QP)} - \mathbf{w}^{(t-1)}$ as a direction in which $\phi$ tends to decrease, and determine the appropriate step size by conducting a line search. In particular, we present the revised algorithm in Algorithm 2.

\begin{center}
\begin{table*}[htb]
\hfill{}
\begin{tabular}{ l }
\hline
{\bf Algorithm 2}:  \\
\hline
\vspace{-3mm} \\
1. Initialize $\mathbf{w}^{\left(0\right)}$, $\mathbf{\mathbf{\boldsymbol{\alpha}}}^{\left(0\right)}$
and $b^{(0)}$ with $w_j \in [0,1]$ for $1 \le j \le p$. \\
\vspace{-3mm} \\
2. The $\boldsymbol{\alpha}$ step: fix $\mathbf{w}^{ (t-1 )}$ and $b^{(t-1)}$, and find
$\balpha^{(t)} = \argmin_{\balpha} \phi( \balpha, b^{(t-1)} ,\mathbf{w}^{ (t-1 )})$. \\
\hspace{4mm} The optimization problem is convex, and independent of the $\lambda_2\|\mathbf{w}\|_1$ term in (\ref{eq:main}). \\
\vspace{-3mm} \\
3. The $b$ step: fix $\mathbf{w}^{ (t-1 )}$ and $\balpha^{(t)}$, and find \\ \hspace{4mm}  $b^{(t)} = \argmin_b \sum_{i=1}^{n} L\big\{y_{i}, \sum_{j=1}^{n}K_{\mathbf{w}^{(t-1)}}(\bx_{i},\bx_{j})\alpha_{j}^{(t)}+b\big\}$. This is a convex \\ \hspace{4mm} optimization with one parameter, and can be solved by standard methods. \\
\vspace{-3mm} \\
4. The $\mathbf{w}$ step: fix $b^{ (t )}$ and $\balpha^{(t)}$, and define $\mathbf{w}^{(\textrm{temp})} = \mathbf{w}^{(t-1)}$. \\
\vspace{-3mm} \\
\hspace{4mm} (a) Define $\bc_{\mathbf{w}^{(\textrm{temp})}}(\mathbf{w}) = A(\mathbf{w}^{(\textrm{temp})}) \mathbf{w} + B(\mathbf{w}^{(\textrm{temp})}) \balpha^{(t)}$. \\
\hspace{10.5mm} Let $\{ \bc_{\mathbf{w}^{(\textrm{temp})}}(\mathbf{w}) \}_i$ be the $i$th element of $\bc_{\mathbf{w}^{(\textrm{temp})}}(\mathbf{w})$. Under the constraint $\mathbf{w} \in [0,1]^p$,\\\hspace{10.5mm} find \\
\hspace{10.5mm}  $\mathbf{w}^{(QP)} = \argmin_{\mathbf{w}} \frac{1}{n} \sum_{i=1}^n  L [y_i, \{ \bc_{\mathbf{w}^{(\textrm{temp})}}(\mathbf{w}) \}_i + b^{(t)}] + \lambda_2 \|\mathbf{w}\|_1 + \lambda_3 \mathbf{w}^T A(\mathbf{w}^{(\textrm{temp})}) \balpha^{(t)} $. \\
\hspace{4mm} (b) Define $\Delta \mathbf{w} = \mathbf{w}^{(QP)} - \mathbf{w}^{(\textrm{temp})}$. Find the best step size $s$ by \\
\vspace{-3mm} \\
\hspace{10.5mm} $s = \argmin_{u \ge 0} \phi (\balpha^{(t)} , b^{ (t )}, \mathbf{w}^{(\textrm{temp})} + u \Delta \mathbf{w})$. \\
\vspace{-3mm} \\
\hspace{4mm} (c) Set $\mathbf{w}^{(\textrm{temp})} = \mathbf{w}^{(\textrm{temp})} + s \Delta \mathbf{w}$. \\
\vspace{-3mm} \\
\hspace{4mm} (d) Repeat steps (a)-(c) until convergence, and set $\mathbf{w}^{(t)} = \mathbf{w}^{(\textrm{temp})}$. \\
\vspace{-3mm} \\
5. Repeat steps 2-4 until convergence. \\
\vspace{-3mm} \\
\hline
\end{tabular}
\hfill{}
\label{algorithm:2}
\end{table*}
\end{center}

In Algorithm 2, one can verify that after updating the parameters, the $\phi$ function value would not increase. This helps to guarantee that we can obtain a stationary point of the objective function using Algorithm 2. In particular, we have the following theorem.

\begin{theorem}
Suppose that the loss function $L$ in (\ref{eq:main}) is a convex and continuously differentiable function, and the variable weighted kernel $K_{\mathbf{w}}$ is a convex or concave and continuously differentiable function of $\mathbf{w}$. Then the solution from Algorithm 2 is a stationary point of the objective function.
\label{thm:algorithm}
\end{theorem}

\noindent {\bf Remark 1}: Theorem~\ref{thm:algorithm} is valid for many loss functions, e.g., the squared error loss in standard regression, and the deviance loss in logistic regression. For many other loss functions that are not differentiable, such as the hinge loss in SVM, or the check loss function in quantile regression, one can consider an alternative continuous approximation to the loss function. For example, \citet{wang_hybrid_2007} proposed the hybrid huberized hinge loss for SVM. One can verify that the hybrid huberized loss meets the condition in Theorem~\ref{thm:algorithm}, and the corresponding solution is a stationary point. Moreover, for many commonly used kernel functions, the assumptions on $K_{\mathbf{w}}$ in Theorem~\ref{thm:algorithm} are satisfied. For example, one can verify that the variable weighted kernel introduced by the Laplacian RKHS, or by the linear kernel when all elements in $\bx$ are non-negative, is convex with respect to $\mathbf{w}$.

\noindent {\bf Remark 2}: Algorithm 2 replaces the quadratic programming step in Algorithm 1 by the descent direction and line search method. This approach is guaranteed to decrease the objective function value at each iteration step, at the cost of a more complex computation. On the other hand, our numerical experience shows that  Algorithm 1 almost always decreases the objective for commonly used kernels and loss functions. Therefore, we use Algorithm 1 in the numerical examples, whereas in each step we check if the objective function decreases. If not, we then employ the line search approach as in Algorithm 2 instead.

\noindent {\bf Remark 3}: Since the objective function can be non-convex, it is possible that the numerical solution is just a stationary point, not the global minimum. To increase the chance of finding the optimal solution, we suggest to use multiple different starting points, compare the corresponding results, and choose the fitted model with the smallest objective function value.

\section{Statistical Learning Theory}
\label{sec:theory}

In this section, we explore the theoretical properties of the proposed DOSK method. In particular, we first study the convergence rate of the excess risk for various learning problems under certain conditions, and then show that DOSK can enjoy selection consistency for high dimensional learning problems. Moreover, we show that the expected loss using the estimated function $\hat{f}$, $E[L\{y,\hat{f}(\bX)\}]$, can be well approximated by the empirical loss on the training data, in the sense that the corresponding difference converges to zero with a fast convergence rate.

To state our theory, we first introduce some technical assumptions, and provide detailed discussions on why these conditions are needed. We also discuss some cases where these conditions are met. We would like to point out that most of the assumptions in this paper are mild and reasonable, which can be satisfied or checked for various real applications.

To begin with, we need to present some further notation. Let $\mathbf{w}^* = (\mathbf{w}_{(1)}^T, \mathbf{w}_{(0)}^T)^T$ be the underlying variable weight vector, where elements in $\mathbf{w}_{(1)}$ are non-zero, and elements in $\mathbf{w}_{(0)}$ are zero. In other words, the predictors in $\bx$ that correspond to $\mathbf{w}_{(0)}$ are noise covariates. Accordingly, one can define $\bx = (\bx_{(1)}^T,\bx_{(0)}^T)^T$, such that predictors in $\bx_{(1)}$ contain useful information for the learning problem. In this paper, we focus on the case that the number of useful predictors is finite (i.e., $|\mathbf{w}_{(1)}| < \infty$). Furthermore, with a little abuse of notation, we let $\|f\|_{\mH} = \|\tilde{f}\|_{\mH}$, where $\tilde{f}$ is the projection of $f$ onto $\mH$.

We impose our first assumption on the distribution of $\bX$ and $\bX_{(1)}$, where $\bX$ and $\bX_{(1)}$ correspond to the $p$ dimension random vector and the vector containing important variables.

\noindent {\bf Assumption 1}: Every element in $\bX$ ranges in $[0,1]$. Furthermore, the distribution of $\bX_{(1)}$ is absolutely continuous with respect to the Lebesgue measure, where the corresponding Radon-Nikodym derivative is bounded away from $0$.

In Assumption 1, we restrict our consideration on $\bX \in [0,1]^p$. One can verify that our theory can be naturally generalized to the case where the elements in $\bX$ are uniformly bounded. We defer the discussion on the second part of Assumption 1 until after Assumption 4.

In the next assumption, we impose some constraints on the kernel function $K(\cdot,\cdot)$.

\noindent {\bf Assumption 2}: The kernel function $K(\cdot,\cdot)$ is separable and $\sup K(\cdot,\cdot) < \infty$. Furthermore, the kernel function $K_{\mathbf{w}^*}(\bx,\cdot)$ is Lipshcitz with respect to $\bx_{(1)}$, i.e. the useful variables vector, in terms of the $L_2$ norm.

The first part of Assumption 2 is very mild, and has been frequently used in the literature. See, for example, \citet{Steinwart2007}, \cite{Blanchard2008}, \citet{zhang_quantile_2015}, among others. It suggests that the corresponding RKHS $\mH$ is not too complex, in the sense that its diameter would not be infinity. The second part is used to ensure that the best learning function using $n$ observations can converge to the underlying function in a fast rate. See the proof of Lemma~\ref{lemma:02} for more details. This assumption is valid for many commonly used kernel functions such as the Gaussian kernel and the polynomial kernel.

In Assumption 3, we assume that $L$ can be treated as a univariate function. This is a very mild condition, and is valid for many learning problems. For example, in standard least squares regression, we have $L(u)=u^2$ where $u=(f-y)$, and in logistic regression, $L(u) = \log\{1+\exp(-u)\}$ where $u=yf$ and $y \in \{+1, -1\}$.

\noindent {\bf Assumption 3}: The loss function $L(u)$ has a second order derivative with $0 < L''(u) < \infty$ for every $u$.

Assumption 3 is needed to ensure that the expected loss function is strictly convex around the underlying optimal solution. Moreover, the second order differentiability helps to control the convergence rate of the estimated function $\hat{f}$ to the best function. See the discussion of Assumption 5 for more details.

Next, we consider assumptions on the function $f(\bx)$. Recall that the learning goal is to obtain $\hat{f}(\bx)$ from the training data set for good prediction performance. Therefore, we consider the ``best" function $f_0$, in the sense that its corresponding expected loss $E [L\{Y, f_0(\bX)\}]$ is the minimum among all possible $E [L\{Y, f (\bX)\}]$. Consequently, $f_0$ can have the best prediction performance under mild conditions. For instance, in classification, $f_0$ can achieve the minimal classification error rate, given that the loss function $L$ is Fisher consistent \citep{Liu07}. We will prove that under certain conditions on $f_0$, the estimated function $\hat{f}$ would converge to $f_0$ with a desirable convergence rate.

\noindent {\bf Assumption 4}: The underlying function $f_0$ has a sparse representation in the RKHS. In particular, there exist $\gamma_1,\ldots,\gamma_m$, $\bz_1,\ldots,\bz_m$, and $b_0$ such that $f_0(\bx) = \sum_{j=1}^m \gamma_j K_{\mathbf{w}^*}(\bz_j, \bx) + b_0$. Here $m$ is a fixed integer, $\gamma_j \ne 0$,  and $\bz_j \in [0,1]^p$ for $j = 1,\ldots, m $.

As a remark, we note that some RKHSs are very rich, in the sense that many functions can be well approximated by $f \in \mH$. For example, \citet{Steinwart2007} proved that all step functions can be approximated by $f$ in the Gaussian RKHS arbitrarily well under mild conditions, and this result can be generalized to the case of continuous functions. However, if $f_0$ does not have a sparse representation in the RKHS, the function in $\mH$ that approximates $f_0$ well may have an infinite norm. When $\hat{f}$ approaches $f_0$ as $n\rightarrow \infty$, $\|\hat{f}\|_{\mH}$ would be unbounded. Consequently, the variation of $\hat{f}$ due to the randomness of the sample can be very large. In the literature, \citet{Bartlett2005}, among others, pointed out that large variation of $\hat{f}$ can lead to suboptimal prediction performance. Assumption 4 ensures that the underlying function $f_0$ has a finite norm in the RKHS. In the proof of Theorem~\ref{thm:f_rate}, we show that with an appropriate $\lambda_1$, the data selection can provide a sparsely represented function $\hat{f}$ whose norm can be bounded away from infinity. This is crucial to prove the convergence of $\hat{f}$ to $f_0$, which further leads to the selection consistency of our DOSK method.

The next assumption ensures that in the updating scheme, $\hat{f}$ would converge to the global solution, once we are at a point that is close enough. To state this assumption, we first introduce some further notation. Define $\|\cdot\|_{*,2}$ to be the restricted $L_2$ norm with respect to the partition of $\mathbf{w}$. In particular, $\| \bx-\bz \|_{*,2} = \|\bx_{(1)} - \bz_{(1)}\|_2$. For any $n \gg m$, we define $(\balpha_n^*, b_n^*)$ as follows. Notice that the empirical loss function value does not change if we switch the order of the pairs $(\bx_i,y_i)$ and $(\bx_j,y_j)$ for $i \ne j$. Hence, without loss of generality, we can assume that $\bx_j$ is the observation that is closest to $\bz_j$ in terms of the $\|\cdot\|_{*,2}$ norm among the training data set $\{ (\bx_i,y_i); \ i=1,\ldots,n \}$, for $j=1,\ldots,m$. When $n \gg m$, we can assume that each $\bx_j$ is distinct (in other words, $\bx_j$ would not be closest to $\bz_u$ and $\bz_v$ simultaneously, compared to other observations). Next, define $(\balpha_n^*, b_n^*)$ such that $\balpha_n^* = (\gamma_1,\ldots,\gamma_m,0,\ldots,0)^T$ with length $n$, $b_n^* = b_0$, and let $f_{\balpha_n^*, b_n^*} (\bx) = \sum_{i=1}^n \alpha_j^* K_{\mathbf{w}^*}(\bx_i,\bx) + b_n^*$. The definition of $(\balpha_n^*, b_n^*)$ helps to show that the approximation error of the DOSK method under Assumption 4 converges to 0 very quickly. See the proof of Lemma~\ref{lemma:04} in the appendix for more discussions.

Before stating Assumption 5, we would like to discuss the second part of Assumption 1, which ensures that with large enough $n$, the underlying function can be well approximated by the sparsely represented function $f_{\balpha_n^*, b_n^*} (\bx)$ from our training data. In particular, Assumption 1 guarantees that as $n \rightarrow \infty$, $f_{\balpha_n^*, b_n^*} (\bx)$ can approach $f_0(\bx)$ with a rate very close to $O_P(n^{-1})$ in terms of the $\| \cdot \|_{2}$ norm. See Lemma~\ref{lemma:02} and the corresponding proof for more discussions.

\noindent {\bf Assumption 5}: For any $p$ and $n \gg m$, there exists a neighborhood $\mN$ of $\big( (\mathbf{w}^*)^T , (\balpha_n^*)^T, b_n^* \big)^T$, such that in $\mN$, the expected loss function $E \left[ \sum_{i=1}^n L \{Y_i, f (\bX_i)\} \right]$ is strictly convex with respect to $( \mathbf{w}^T, \balpha^T, b )^T$.

Assumption 5 is necessary for our theory, because if the loss function is not strictly convex, a small perturbation in the training data set can lead to a significant change of $\hat{f}$. See, for example, the discussion on a similar issue for quantile regression using the check loss function in \citet{Li2008}. Consequently, the convergence rate of $\hat{f}$ to $f_0$ can be difficult to obtain. To our knowledge, there has been no theoretical result on selection consistency that does not rely on the assumption or fact of local convexity. Notice that Assumption 3 is important to the validity of Assumption 5, because if $L$ is not strictly convex, it is likely that the expected loss function is not convex even if the kernel function is locally convex. For instance, if we use the hinge loss $L(u) = [1-u]_+$ which is piecewise linear, Assumption 5 cannot be satisfied.

Next, we impose constraints on the signal strength in the learning problem. For variables weighted learning, the $j$th predictor provides useful information if and only if the weight $w_j$ is positive. Variable selection consistency means that $\textrm{sign}(\hat{w}_j) = \textrm{sign}(w_j)$ for all $j$ with a high probability, where $\textrm{sign}(0) = 0$. The next assumption is an important part of sufficient conditions for variable selection consistency.

\noindent {\bf Assumption 6}: For any $w_j$ in $\mathbf{w}_{(1)}$, $\frac{\partial E [L\{Y, f_0(\bX)\}] }{\partial w_j} \mid_{w_j=0, \ w_i=w^*_i, \ i \ne j} < 0$, and for any $w_j$ in $\mathbf{w}_{(0)}$, $\frac{\partial E [L\{Y, f_0(\bX)\}] }{\partial w_j} \mid_{w_j=0, \ w_i=w^*_i, \ i \ne j} \ge 0$. Here $w^*_i$ is the $i$th element of $\mathbf{w}^*$.

In Assumption 6, we measure the signal strength of $w_j$ by its partial derivative with respect to the expected loss function evaluated at $\mathbf{w}^*$ (except the $j$th weight is at zero). In the literature, there are many existing assumptions on the signal strength that are (essentially) similar to Assumption 6. For example, one can verify that for regular linear regression with the squared error loss, Assumption 6 reduces to that the non-zero coefficients are bounded away from zero. This is analogous to the assumptions considered in \citet{fan2004nonconcave} and \citet{fan2010selective}, among others. Furthermore, we require the partial derivative with respect to the noise covariates are non-negative.

In the last assumption, we focus on regression problems, where $Y = f_0(\bX) + \epsilon(\bX)$ with $\epsilon(\bX)$ being the random error term. Notice that we include both the homoscedastic and the heteroscedastic cases here, as $\epsilon$ can have different distributions for different $\bX$. If the distribution of $\epsilon$ has a very heavy tail, there is a large probability that we observe a $y_i$ that is very far away from $f_0 (\bx_i)$. This outlier can lead to a severely biased estimation $\hat{f}$. Assumption 7 aims to control the probability of an extreme $y_i$, which can help to bound the magnitude of the estimated $\hat{b}$. Recall that if a random variable $U$ is sub-Gaussian with parameter $s$, then $\textrm{pr}(|U|>u) \le 2\exp(-u^2/s)$ for large enough $u$.

\noindent {\bf Assumption 7}: In a regression problem, the error term $\epsilon(\bX)$ follows a sub-Gaussian distribution with a universal parameter $s < \infty$ for any $\bX$.

Assumption 7 is very general, as many distributions are sub-Gaussian. For example, in linear regression, we often assume that $\epsilon \sim N(0,\sigma^2)$ with finite $\sigma$. This is a homoscedastic case of Assumption 7, and normal random variables are known to be sub-Gaussian. Furthermore, all random variables with bounded ranges are sub-Gaussian, and distributions with small kurtosis are sub-Gaussian.

We are ready to present our main theorems. The first theorem studies the convergence rate of $\hat{f}$ to $f_0$. Recall that $a \vee b = \max(a,b)$ for $a,b\in \mbbR$.

\begin{theorem}
Suppose Assumptions 1-7 hold, and $\log(p)/\sqrt{n} \rightarrow 0$ as $n\rightarrow \infty$. If we choose $\lambda_1 = O\{\log(n)^{-1}\}$, $\lambda_2 = O [ \{\log(p) \vee \log(n)\} / \sqrt{n} ]$, and $\lambda_3 = o(\lambda_1)$ in (\ref{eq:main}), we have that the corresponding global solution $(\hat{\mathbf{w}}^T,\hat{\balpha}^T,\hat{b})^T$ to (\ref{eq:main}) satisfies that $\| \hat{f} - f_0 \|_2 = O_P\{ \log(n) / \sqrt{n} \} $, where $\hat{f}(\bx) = \sum_{j=1}^n \hat{\alpha}_j K_{\hat{\mathbf{w}}} (\bx,\bx_j)+ \hat{b}$.
\label{thm:f_rate}
\end{theorem}

Theorem~\ref{thm:f_rate} suggests that $\hat{f}$ converges to $f_0$ at a rate very close to the ``parametric rate" $O_P(n^{-1/2})$. Comparing Theorem~\ref{thm:f_rate} with the theoretical results in \citet{zhang_quantile_2015}, one can see that the multiple penalties in (\ref{eq:main}) do not affect the performance of $\hat{f}$, as long as the corresponding $\lambda$'s are appropriately selected. This helps to justify that our DOSK method can avoid the issue of over-penalization by carefully choosing the tuning parameters.

Next, we study the selection consistency of our DOSK method. Our results suggest that we can have selection consistency if $p$ is of a polynomial order of $n$.

\begin{theorem}
Suppose Assumptions 1-7 hold. Furthermore, assume that $\log(p)/\sqrt{n} \rightarrow 0$ as $n\rightarrow \infty$. If we choose $\lambda_1 = O\{\log(n)^{-1}\}$, $\lambda_2 = O [ \{\log(p) \vee \log(n)\} / \sqrt{n} ]$, and $\lambda_3 = o(\lambda_1)$ in (\ref{eq:main}), we have that the corresponding global solution $(\hat{\mathbf{w}}^T,\hat{\balpha}^T,\hat{b})^T$ to (\ref{eq:main}) satisfies that, with probability tending to 1 as $n\rightarrow \infty$, $\textrm{sign}(\hat{w}_j) = \textrm{sign}(w_j^*)$ for $j=1,\ldots,p$, where $w_j^*$ is the $j$th element of $\mathbf{w}^*$.
\label{thm:sele_con}
\end{theorem}

Theorem~\ref{thm:sele_con} shows that our DOSK method can enjoy the desirable asymptotic selection consistency at the global solution. In other words, if the sample size is large, one can often correctly identify the important and unimportant variables in the learning problem. This can help researchers to obtain a better understanding of the relationship between predictors and the response, and provide a more interpretable model for future prediction.

The next theorem studies the prediction performance of the obtained $\hat{f}$. In particular, since one uses the loss function $L$ to measure the goodness of fit of $\hat{f}$, it is desirable to obtain a bound for the expected loss $E[L \{Y, \hat{f}(\bX)\}]$. For example, in regression problems, $E[L \{Y, \hat{f}(\bX)\}]$ indicates the average prediction error using $\hat{f}$. In margin-based classification where the loss function $L$ dominates the $0-1$ loss function (which is further equivalent to the prediction error rate), $E[L \{Y, \hat{f}(\bX)\}]$ can be regarded as an upper bound of the future misclassification rate. In the next theorem, we show that under the assumptions specified above, the empirical measurement $n^{-1} \sum_{i=1}^n [L \{y_i, \hat{f}(\bx_i)\}]$ converges to its expectation $E[L \{Y, \hat{f}(\bX)\}]$ at the rate $O_P [ \{\log(p) \vee \log(n)\} / \sqrt{n} ]$.

\begin{theorem}
Suppose Assumptions 1-7 hold. Furthermore, assume that $\log(p)/\sqrt{n} \rightarrow 0$ as $n\rightarrow \infty$. If we choose $\lambda_1 = O\{\log(n)^{-1}\}$, $\lambda_2 = O [ \{\log(p) \vee \log(n)\} / \sqrt{n} ]$, and $\lambda_3 = o(\lambda_1)$ in (\ref{eq:main}), we have that the corresponding global solution $(\hat{\mathbf{w}}^T,\hat{\balpha}^T,\hat{b})^T$ to (\ref{eq:main}) satisfies that, $ | E[L \{Y, \hat{f}(\bX)\}] - n^{-1} \sum_{i=1}^n [L \{y_i, \hat{f}(\bx_i)\}] | = O_P [ \{\log(p) \vee \log(n)\} / \sqrt{n} ]$, where $\hat{f}(\bx) = \sum_{j=1}^n \hat{\alpha}_j K_{\hat{\mathbf{w}}} (\bx,\bx_j)+ \hat{b}$.
\label{thm:risk_bnd}
\end{theorem}

Theorem~\ref{thm:risk_bnd} shows that the empirical average loss $n^{-1} \sum_{i=1}^n [L \{y_i, \hat{f}(\bx_i)\}]$ from the training data set, can be a good estimate of the expected loss $E[L \{Y, \hat{f}(\bX)\}]$. As discussed above, this empirical loss can provide valuable information on the prediction performance of $\hat{f}$.

As a remark, we would like to point out that our theorems can be generalized to the case of local solutions, provided that similar conditions as in Assumptions 4-6 are met. For example, the convexity of local solutions can be stated in an analogous manner as in Assumption 5, and the corresponding signal strength can be measured by the partial derivatives as in Assumption 6.

\section{Numerical Analysis}
\label{sec:numerical}

In this section, we use regression and classification as examples of learning techniques, and explore the numerical performance of our proposed DOSK method using simulated and real data sets. In Section~\ref{sec:simu}, we study the empirical prediction behavior of DOSK using synthetic data sets, and in Section~\ref{sec:real}, we examine the performance of DOSK in real data applications. We compare our method with some existing approaches in the literature. In particular, for regression problems, we compare our DOSK method with the standard linear ridge regression, LASSO, standard $L_2$ kernel learning as in (\ref{eq:generalL2new}), COSSO and KNIFE. Moreover, we implement the Sure Independence Screening (SIS) and Recursive Feature Elimination (RFE) methods with $L_2$ kernel learning. Notice here the generalization of SIS from linear learning to kernel learning is analogous to the approach discussed in \citet{guyon_gene_2002}. We employ the squared error loss function for all regression techniques. For classification methods, we use the SVM hinge loss for DOSK, and compare with the standard kernel SVM, kernel SIS SVM, kernel RFE SVM and KNIFE SVM.

In all numerical examples, we select the tuning parameters as follows. For our DOSK method, because there are three tuning parameters $\lambda_1$-$\lambda_3$ and potential kernel parameters (such as the $\gamma$ parameter in the Gaussian kernel), we fix $\lambda_{3}=0.5$, and let other parameters be selected from a set of candidates. In particular, we let $\lambda_1$ vary in $\{ 0,0.25,0.5 \}$, and let $\lambda_2$ vary in $\{2^i; \ i=-3,-2,\ldots,2,3\}$. As we will show in Section 4.1 that the selection of $\lambda_{3}$, the tuning parameter for the quadratic kernel regularization term, does not appear to play an essential role in maximizing the prediction accuracy of DOSK as long as its value is taken within a certain range. For the kernel parameters, because we use the Gaussian and Laplacian kernels (whose kernel functions are discussed in Section~\ref{sec:dosk}) in our analysis, we let the parameter $\gamma$ vary in $\{0.1, 0.2, \ldots, 0.9, 1\}$, a candidate set whose range always covers $1/2\hat{\sigma}^2$ where $\hat{\sigma}$ is the median of the Euclidean distances between each pair of the observations. In our experience, this tuning procedure works reasonably well for the numerical examples in this paper. For real applications, one can perform finer tuning procedures using a larger candidate set of tuning parameters. For other existing approaches except SIS and RFE, the tuning parameters are chosen in an analogous manner. The best set of tuning parameters that minimizes the prediction error in five fold cross validations on the training data set is then selected, and we report the corresponding prediction errors on a separate testing data set. Here the prediction error for regression examples is measured by the Mean Prediction Error \citep[MPE,][]{hastie_elements_2011}, $\frac{1}{n}\sum_{i=1}^{n}\{ \hat{f}(\bx_i) -y_i \}^2$. The error measure for classification problems is the misclassification rate (MCR), $\frac{1}{n} \sum_{i=1}^n I [y_i \ne \textrm{sign}\{\hat{f}(\bx_i)\} ]  $, where $I(\cdot)$ is the indicator function.

\subsection{Simulated Examples}
\label{sec:simu}

In this section, we conduct four simulated examples to demonstrate the performance of our DOSK method. The first two examples are regression problems, and the last two are classification problems. In each example, we let the responses depend only on several predictors, and we add noise covariates in the date sets. We denote by $p_0$ the number of noise predictors. To assess various methods, we repeat each example 50 times and report the average prediction errors on the training and testing data sets. Furthermore, for all the methods that have variable selection, we report the True Positive (TP) rates and False Negative (FN) rates of predictors to compare the corresponding performance on variable selection.

\noindent {\bf Regression Example 1:} For this example, the response depends only on one predictor. In particular, we have $y_{i} = 10 \sin (x_{i1}) I(0<x_{i1}<2\pi) + \epsilon_{i}$ where $x_{i1}$ is the first predictor of the $i$th observation. Here $x_{ij}$ follows a uniform distribution within $[-2\pi,4\pi]$ for $j=1,\cdots,1+p_0$, and the error term $\epsilon$ is generated from the standard normal distribution. In this example, we let $p_0=2$ and $p_0=8$, and choose the size of the training data set to be $50$ and $100$. The size of the testing set is 10 times larger than that of the training set. We use the Laplacian kernel in this example.

\begin{figure}[htb]
\begin{center}
\includegraphics[width=0.6\textwidth,totalheight=0.6\textwidth]{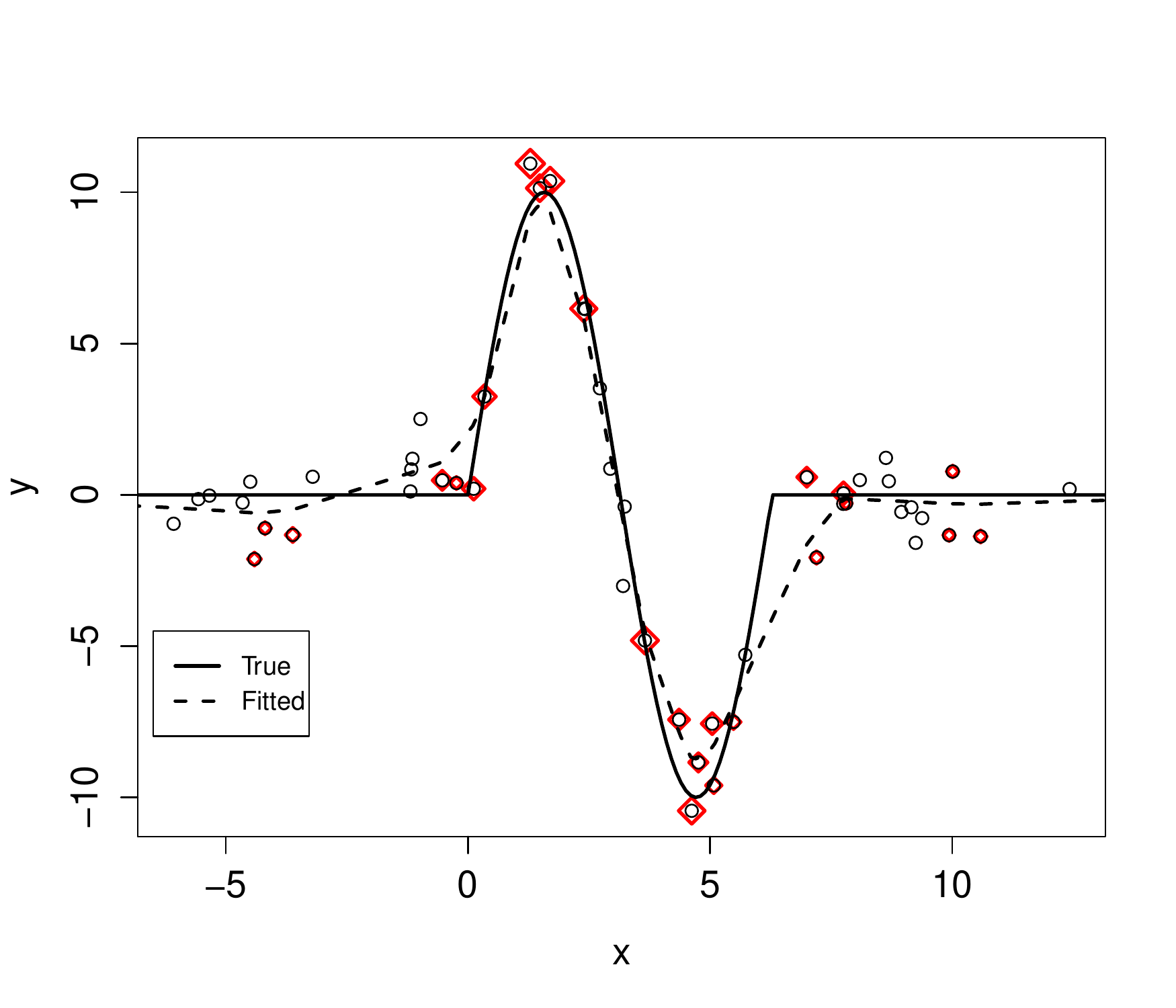}
\end{center}
\caption{Plot of the underlying $f_0$ (solid) and fitted $\hat{f}$ by DOSK (dashed) when $n=100$ and $p_0=2$. Observations with non-zero $\hat{\alpha}_j$'s are highlighted in red. One can see that the data sparsity penalty tends to choose observations that are closer to $0$, $\pi/2$, $3\pi/2$ and $2\pi$ for the function representation.}
\label{simufig1}
\end{figure}

The numerical results for Regression Example 1 are reported in Table~\ref{simutab1}. One can see that the ridge regression and LASSO perform poorly using linear learning, as the underlying function $f_0$ is highly nonlinear. Note that the standard kernel learning method with the $L_2$ penalty has very small prediction error rate on the training data set. This shows that the corresponding models can fit the training observations very well. However, the errors on the testing data set are very large. This suggests that without appropriate variable selection, the performance of standard kernel learning can be greatly undermined by overfitting. Moreover, the SIS and RFE approaches can also have overfitting issues, which are partly due to their large FN rates. Compared to these methods, KNIFE and our DOSK work competitively. Note that the prediction error of COSSO is also good with a large sample size ($n=100$). However, the corresponding variation is significantly larger than that of KNIFE or DOSK. This suggests that decomposing the nonlinear function into a sum of orthogonal components can be instable for some kernels. Furthermore, as the underlying function can be well approximated by functions that have sparse presentations, our DOSK method works better than KNIFE. This is similar to the findings in \citet{zhang_quantile_2015}. To demonstrate the effect of data selection, in Figure~\ref{simufig1}, we plot the fitted regression function $\hat{f}$ from our DOSK method in a typical replicate, and the underlying function $f_0$ as a comparison. Moreover, we plot all the training observations, and highlight the selected ones, whose corresponding $\hat{\alpha}_j$'s are non-zero. One can see that because we are using the Laplacian kernel which has a singularity at $0$ and smooth elsewhere, the data sparsity penalty tends to choose the observations that are closer to the ``sharp turns" of $f_0$ for representation. This helps to build a model that is smooth when the curvature of $f_0$ is small, thus prevents overfitting from using all observations in the kernel function representation.

\begin{center}
\begin{table*}[htb]
{\scriptsize{}}%
\hfill{} \begin{tabular}{c|c|cccc|cccc}
\hline
\hline
\multirow{2}{*}{{\scriptsize{}$p_0$ }} & \multirow{2}{*}{{\scriptsize{}Method}} & \multicolumn{4}{c|}{{\scriptsize{}$n=50$}} & \multicolumn{4}{c}{{\scriptsize{}$n=100$}}\tabularnewline
\cline{3-10}
 &  & {\scriptsize{}Train MPE} & {\scriptsize{}Test MPE} & {\scriptsize{}TP} & {\scriptsize{}FN} & {\scriptsize{}Train MPE} & {\scriptsize{}Test MPE} & {\scriptsize{}TP} & {\scriptsize{}FN}\tabularnewline
\hline
\multirow{8}{*}{{\scriptsize{}2}} & {\scriptsize{}Linear Ridge} & {\scriptsize{}15.89 (4.46)} & {\scriptsize{}17.96 (1.33)} & {\scriptsize{}-} & {\scriptsize{}-} & {\scriptsize{}16.29 (3.46)} & {\scriptsize{}17.82 (1.16)} & {\scriptsize{}-} & {\scriptsize{}-}\tabularnewline
\cline{2-10}
 & {\scriptsize{}LASSO} & {\scriptsize{}15.89 (4.47)} & {\scriptsize{}17.96 (1.32)} & {\scriptsize{}1} & {\scriptsize{}0.49} & {\scriptsize{}16.29 (3.46)} & {\scriptsize{}17.82 (1.17)} & {\bf\scriptsize{}1} & {\scriptsize{}0.5}\tabularnewline
\cline{2-10}
 & {\scriptsize{}$L_2$ Kernel} & {\scriptsize{}2.06 (0.45)} & {\scriptsize{}11.17 (2.00)} & {\scriptsize{}-} & {\scriptsize{}-} & {\scriptsize{}2.09 (0.38)} & {\scriptsize{}7.36 (1.55)} & {\scriptsize{}-} & {\scriptsize{}-}\tabularnewline
\cline{2-10}
 & {\scriptsize{}SIS} & {\scriptsize{}8.22 (5.50)} & {\scriptsize{}12.20 (7.13)} & {\scriptsize{}0.42} & {\scriptsize{}0.29} & {\scriptsize{}5.39 (5.85)} & {\scriptsize{}7.54 (7.51)} & {\scriptsize{}0.68} & {\scriptsize{}0.16}\tabularnewline
\cline{2-10}
 & {\scriptsize{}RFE} & {\scriptsize{}4.77 (3.91)} & {\scriptsize{}10.57 (6.05)} & {\scriptsize{}0.44} & {\scriptsize{}0.30} & {\scriptsize{}3.10 (3.51)} & {\scriptsize{}5.44 (5.02)} & {\scriptsize{}0.7} & {\scriptsize{}0.16}\tabularnewline
\cline{2-10}
 & {\scriptsize{}COSSO} & {\scriptsize{}7.05 (6.56)} & {\scriptsize{}11.99 (10.32)} & {\scriptsize{}0.56} & {\scriptsize{}0.39} & {\scriptsize{}0.96 (1.29)} & {\scriptsize{}1.99 (2.58)} & {\scriptsize{}0.98} & {\scriptsize{}0.53}\tabularnewline
\cline{2-10}
 & {\scriptsize{}KNIFE} & {\scriptsize{}3.66 (0.48)} & {\scriptsize{}6.14 (2.00)} & {\bf\scriptsize{}1} & {\scriptsize{}0.14} & {\scriptsize{}2.35 (0.19)} & {\scriptsize{}3.03 (0.57)} & {\bf\scriptsize{}1} & {\bf  \scriptsize{}0}\tabularnewline
\cline{2-10}
 & {\scriptsize{}DOSK} & {\scriptsize{}1.42 (0.21)} & \textbf{\scriptsize{}3.40 (2.92)} & \textbf{\scriptsize{}1} & \textbf{\scriptsize{}0.04} & {\scriptsize{}0.92 (0.13)} & \textbf{\scriptsize{}1.42 (0.19)} & \textbf{\scriptsize{}1} & \textbf{\scriptsize{}0}\tabularnewline
\hline
\multirow{8}{*}{{\scriptsize{}8}} & {\scriptsize{}Linear Ridge} & {\scriptsize{}13.77 (2.89)} & {\scriptsize{}18.09 (1.55)} & {\scriptsize{}-} & {\scriptsize{}-} & {\scriptsize{}16.11 (2.78)} & {\scriptsize{}17.68 (1.03)} & {\scriptsize{}-} & {\scriptsize{}-}\tabularnewline
\cline{2-10}
 & {\scriptsize{}LASSO} & {\scriptsize{}13.77 (2.89)} & {\scriptsize{}18.12 (2.15)} & {\bf\scriptsize{}1} & {\scriptsize{}0.87} & {\scriptsize{}16.13 (2.77)} & {\scriptsize{}17.61 (1.02)} & {\bf\scriptsize{}1} & {\scriptsize{}0.88}\tabularnewline
\cline{2-10}
 & {\scriptsize{}$L_2$ Kernel} & {\scriptsize{}0.05 (0.01)} & {\scriptsize{}17.26 (1.52)} & {\scriptsize{}-} & {\scriptsize{}-} & {\scriptsize{}0.05 (0.01)} & {\scriptsize{}15.76 (1.05)} & {\scriptsize{}-} & {\scriptsize{}-}\tabularnewline
\cline{2-10}
 & {\scriptsize{}SIS} & {\scriptsize{}3.94 (2.04)} & {\scriptsize{}16.18 (4.44)} & {\scriptsize{}0.46} & {\scriptsize{}0.31} & {\scriptsize{}3.07 (1.90)} & {\scriptsize{}9.01 (3.95)} & {\scriptsize{}0.86} & {\scriptsize{}0.26}\tabularnewline
\cline{2-10}
 & {\scriptsize{}RFE} & {\scriptsize{}9.83 (4.97)} & {\scriptsize{}16.18 (12.30)} & {\scriptsize{}0.54} & {\bf\scriptsize{}0.24} & {\scriptsize{}6.44 (5.73)} & {\scriptsize{}10.29 (6.03)} & {\scriptsize{}0.86} & {\scriptsize{}0.25}\tabularnewline
\cline{2-10}
 & {\scriptsize{}COSSO} & {\scriptsize{}12.27 (40.97)} & {\scriptsize{}19.93 (12.30)} & {\scriptsize{}0.54} & \textbf{\scriptsize{}0.24} & {\scriptsize{}6.44 (5.73)} & {\scriptsize{}10.29 (8.66)} & {\scriptsize{}0.76} & {\scriptsize{}0.25}\tabularnewline
\cline{2-10}
 & {\scriptsize{}KNIFE} & {\scriptsize{}2.40 (0.53)} & {\scriptsize{}13.89 (3.64)} & {\bf\scriptsize{}1} & {\scriptsize{}0.42} & {\scriptsize{}1.58 (0.18)} & {\scriptsize{}2.69 (1.99)} & {\textbf{\scriptsize{}1}} & {\scriptsize{}0.22}\tabularnewline
\cline{2-10}
 & {\scriptsize{}DOSK} & {\scriptsize{}2.70 (0.59)} & \textbf{\scriptsize{}10.80 (5.59)} & {\scriptsize{}0.95} & {\scriptsize{}0.29} & {\scriptsize{}1.12 (0.20)} & \textbf{\scriptsize{}2.15 (2.81)} & \textbf{\scriptsize{}1} & \textbf{\scriptsize{}0.20}\tabularnewline
\hline
\hline
\end{tabular} \hfill{}
\caption{Results of Regression Example 1. The numbers in parentheses show
the corresponding standard deviations. MPE stands for mean prediction error,  TP and FN represent true positive rates and false negative rates, respectively.}
\label{simutab1}
\end{table*}
\end{center}

\noindent {\bf Regression Example 2:} In this example, the response $Y$ depends on $4$ predictors. In particular,
\begin{align*}
y_{i} = 10 \sum_{j=1}^4 \exp (-x_{ij}^2) + \epsilon_{i},
\end{align*}
where the error term follows standard normal distribution, and $x_{ij}$ follows a uniform distribution in $[-6,6]$ for $j=1,\ldots,4$. The number of noise covariates and sizes of the training and testing data sets are the same as in Regression Example 1. We use the Gaussian kernel in this example. The prediction performance and variable selection results for Regression Example 2 are reported in Table~\ref{simutab2}, and one can draw similar conclusions as in Regression Example 1.

\begin{center}
\begin{table*}[htb]
{\scriptsize{}}%
\hfill{}
\begin{tabular}{c|c|cccc|cccc}
\hline
\multirow{2}{*}{{\scriptsize{}$p_0$}} & \multirow{2}{*}{{\scriptsize{}Method}} & \multicolumn{4}{c|}{{\scriptsize{}$n=50$}} & \multicolumn{4}{c}{{\scriptsize{}$n=100$}}\tabularnewline
\cline{3-10}
 &  & {\scriptsize{}Train MPE} & {\scriptsize{}Test MPE} & {\scriptsize{}TP} & {\scriptsize{}FN} & {\scriptsize{}Training MPE} & {\scriptsize{}Test MPE} & {\scriptsize{}TP} & {\scriptsize{}FN}\tabularnewline
\hline
\multirow{8}{*}{{\scriptsize{}2}} & {\scriptsize{}Linear Ridge} & {\scriptsize{}30.20 (7.34)} & {\scriptsize{}35.01 (2.54)} & {\scriptsize{}-} & {\scriptsize{}-} & {\scriptsize{}32.11 (5.91)} & {\scriptsize{}33.97 (2.09)} & {\scriptsize{}-} & {\scriptsize{}-}\tabularnewline
\cline{2-10}
 & {\scriptsize{}LASSO} & {\scriptsize{}30.19 (7.34)} & {\scriptsize{}35.00 (2.57)} & {\scriptsize{}0.99} & {\scriptsize{}0.50} & {\scriptsize{}32.10 (5.91)} & {\scriptsize{}33.97 (2.09)} & {\scriptsize{}1} & {\scriptsize{}0.50}\tabularnewline
\cline{2-10}
 & {\scriptsize{}$L_2$ Kernel} & {\scriptsize{}0.05 (0.02)} & {\scriptsize{}28.01 (2.57)} & {\scriptsize{}-} & {\scriptsize{}-} & {\scriptsize{}0.04 (0.01)} & {\scriptsize{}23.94 (2.09)} & {\scriptsize{}-} & {\scriptsize{}-}\tabularnewline
\cline{2-10}
 & {\scriptsize{}SIS} & {\scriptsize{}1.07 (2.09)} & {\scriptsize{}30.92 (3.53)} & {\scriptsize{}0.34} & {\scriptsize{}0.31} & {\scriptsize{}1.92 (2.70)} & {\scriptsize{}29.61 (3.62)} & {\scriptsize{}0.29} & {\scriptsize{}0.41}\tabularnewline
\cline{2-10}
 & {\scriptsize{}RFE} & {\scriptsize{}8.75 (8.22)} & {\scriptsize{}32.15 (4.05)} & {\scriptsize{}0.34} & {\scriptsize{}0.32} & {\scriptsize{}14.32 (9.20)} & {\scriptsize{}30.34 (3.53)} & {\scriptsize{}0.30} & {\scriptsize{}0.27}\tabularnewline
\cline{2-10}
 & {\scriptsize{}COSSO} & {\scriptsize{}14.56 (4.60)} & {\scriptsize{}31.45 (11.10)} & {\scriptsize{}0.49} & \textbf{\scriptsize{}0.17} & {\scriptsize{}16.33 (8.93)} & {\scriptsize{}21.09 (9.62)} & {\scriptsize{}0.48} & {\textbf{\scriptsize{}0.11}}\tabularnewline
\cline{2-10}
 & {\scriptsize{}KNIFE} & {\scriptsize{}6.56 (1.33)} & {\scriptsize{}21.26 (3.12)} & \textbf{\scriptsize{}1} & {\scriptsize{}0.49} & {\scriptsize{}5.99 (0.54)} & {\scriptsize{}12.99 (1.29)} & \textbf{\scriptsize{}1} & {\scriptsize{}0.18}\tabularnewline
\cline{2-10}
 & {\scriptsize{}DOSK} & {\scriptsize{}2.14 (0.61)} & \textbf{\scriptsize{}18.25 (3.70)} & \textbf{\scriptsize{}1} & {\scriptsize{}0.54} & {\scriptsize{}2.60 (0.31)} & \textbf{\scriptsize{}9.86 (1.44)} & \textbf{\scriptsize{}1} & {\scriptsize{}0.12}\tabularnewline
\hline
\multirow{8}{*}{{\scriptsize{}8}} & {\scriptsize{}Linear Ridge} & {\scriptsize{}26.28 (7.09)} & {\scriptsize{}33.95 (3.05)} & {\scriptsize{}-} & {\scriptsize{}-} & {\scriptsize{}30.06 (5.60)} & {\scriptsize{}34.21 (1.73)} & {\scriptsize{}-} & {\scriptsize{}-}\tabularnewline
\cline{2-10}
 & {\scriptsize{}LASSO} & {\scriptsize{}26.26 (7.07)} & {\scriptsize{}33.94 (3.04)} & {\scriptsize{}1} & {\scriptsize{}0.88} & {\scriptsize{}29.06 (5.41)} & {\scriptsize{}33.17 (1.69)} & {\scriptsize{}1} & {\scriptsize{}0.88}\tabularnewline
\cline{2-10}
 & {\scriptsize{}$L_2$ Kernel} & {\scriptsize{}0.05 (0.02)} & {\scriptsize{}33.97 (3.05)} & {\scriptsize{}-} & {\scriptsize{}-} & {\scriptsize{}0.04 (0.01)} & {\scriptsize{}26.23 (1.73)} & {\scriptsize{}-} & {\scriptsize{}-}\tabularnewline
\cline{2-10}
 & {\scriptsize{}SIS} & {\scriptsize{}0.05 (0.03)} & {\scriptsize{}33.63 (2.94)} & {\scriptsize{}0.32} & {\scriptsize{}0.33} & {\scriptsize{}0.04 (0.01)} & {\scriptsize{}33.71 (1.84)} & {\scriptsize{}0.31} & {\scriptsize{}0.35}\tabularnewline
\cline{2-10}
 & {\scriptsize{}RFE} & {\scriptsize{}10.54 (7.79)} & {\scriptsize{}32.90 (3.50)} & {\scriptsize{}0.33} & {\textbf{\scriptsize{}0.18}} & {\scriptsize{}13.92 (10.32)} & {\scriptsize{}32.25 (3.30)} & {\scriptsize{}0.32} & {\scriptsize{}0.19}\tabularnewline
\cline{2-10}
 & {\scriptsize{}COSSO} & {\scriptsize{}18.36 (7.82)} & {\scriptsize{}35.54 (6.68)} & {\scriptsize{}0.31} & {\scriptsize{}0.25} & {\scriptsize{}16.41 (7.13)} & {\scriptsize{}27.14 (7.13)} & {\scriptsize{}0.51} & {\scriptsize{}0.18}\tabularnewline
\cline{2-10}
 & {\scriptsize{}KNIFE} & {\scriptsize{}5.47 (0.78)} & {\scriptsize{}25.53 (4.03)} & \textbf{\scriptsize{}0.99} & {\scriptsize{}0.46} & {\scriptsize{}5.53 (0.50)} & {\scriptsize{}14.52 (2.41)} & \textbf{\scriptsize{}1} & {\scriptsize{}0.17}\tabularnewline
\cline{2-10}
 & {\scriptsize{}DOSK} & {\scriptsize{}1.54 (0.33)} & \textbf{\scriptsize{}23.97 (6.10)} & {\textbf{\scriptsize{}0.99}} & {\scriptsize{}0.36} & {\scriptsize{}2.37 (0.28)} & \textbf{\scriptsize{}10.70 (3.20)} & \textbf{\scriptsize{}1} & \textbf{\scriptsize{}0.15}\tabularnewline
\hline
\end{tabular}
\hfill{} \caption{Results of Regression Example 2. The numbers in parentheses show
the corresponding standard deviations. MPE stands for mean prediction error,  TP and FN represent true positive rates and false negative rates, respectively.}
\label{simutab2}
\end{table*}
\end{center}

\begin{figure}[htb]
\begin{center}
\includegraphics[width=0.6\textwidth,totalheight=0.6\textwidth]{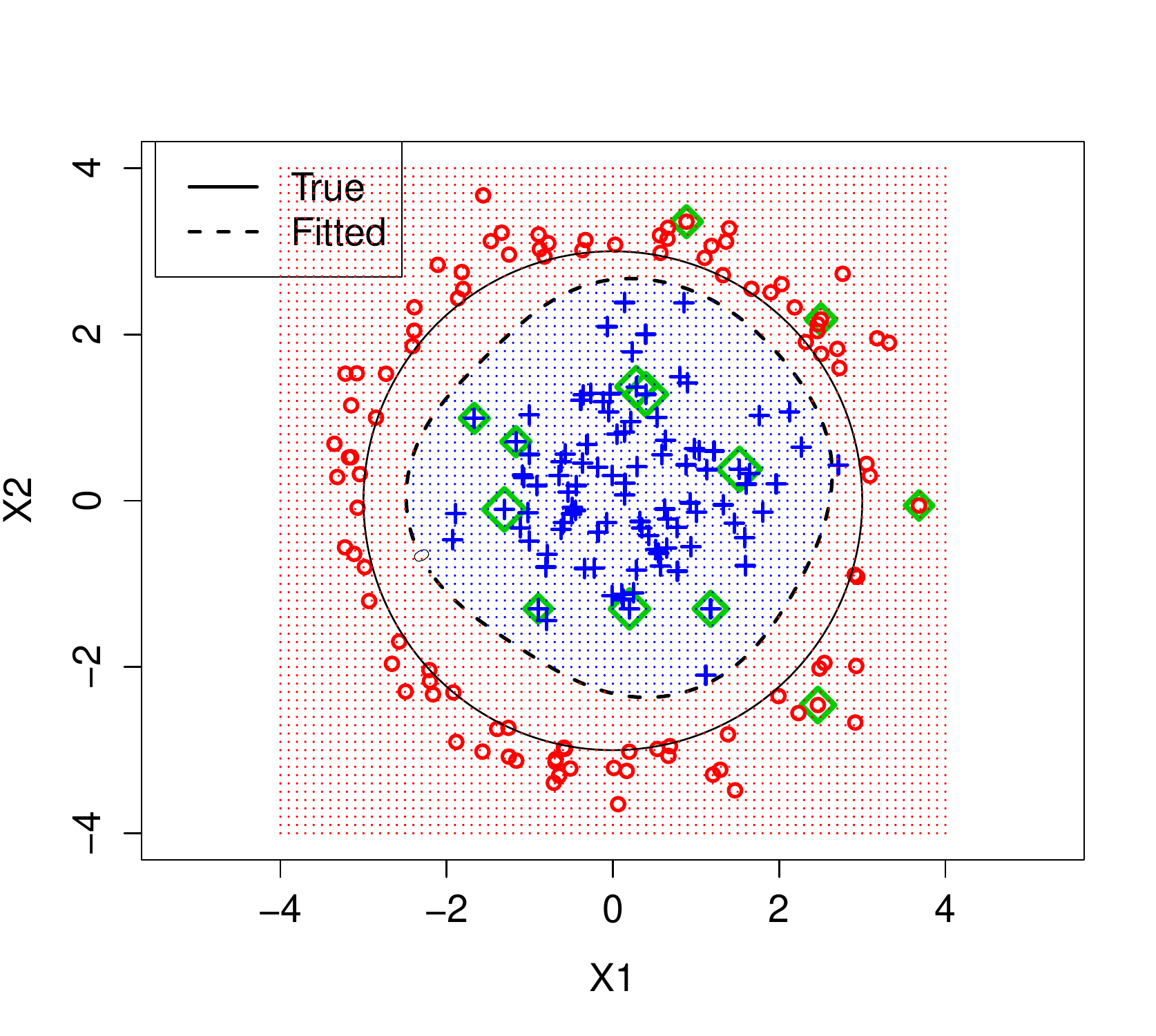}
\end{center}
\caption{Plot of the underlying classification boundary (solid circle) and estimated boundary by DOSK (dashed circle) when $n=200$ and $p_0=8$. Observations with non-zero $\hat{\alpha}_j$'s are highlighted in green.}
\label{simufig2}
\end{figure}

\noindent {\bf Classification Example 1:} In this example, we consider a binary classification problem, where the prior probabilities $\pr(Y=+1) = \pr(Y=-1) = 1/2$. The posterior probabilities $\pr(Y=+1\mid \bX=\bx)$ depend on two predictors. In particular, the distribution of $x_{\cdot 1}$ and $x_{\cdot 2}$ for the first class is $N\{(0,0)^T, I_2\}$, where $x_{\cdot j}$ represents the $j$th predictor, and $I_2$ is the $2 \times 2$ identity matrix. For the second class, the distribution of $x_{\cdot 1}$ and $x_{\cdot 2}$ is proportional to the restricted joint normal distribution $N\{(0,0)^T, I_2\} \mid 9< (x_{\cdot 1}^2 + x_{\cdot 2}^2) <16$. To illustrate the marginal distribution of $x_{\cdot 1}$ and $x_{\cdot 2}$, we plot the first two covariates for a typical sample in Figure~\ref{simufig2}. In this example, we let $p_0=0,4,8$, and add independent noise variables following $N\left(0,0.1\right)$ in the data set. The number of observations in the training data set is $200$, and in the testing $2000$. Note that a similar example was previously used in \citet{hastie_elements_2011}. The Gaussian kernel is used.

The simulation results are reported in Table~\ref{simutab3}. One can see that when there are no noise predictors, all the methods can provide similar classification performance, with our DOSK method being slightly better. When the number of noise covariates increases, the prediction performance of $L_2$ kernel SVM, SIS and RFE deteriorates. On the other hand, the KNIFE method and our DOSK work competitively. Moreover, in this example, the classification boundary $(x_{\cdot 1}^2 + x_{\cdot 2}^2 = 9)$ is relatively simple (see Figure~\ref{simufig2} for an illustration). Hence, functions with sparse representations in the dual space can separate the two classes well. Consequently, our DOSK method works better than the KNIFE approach. In terms of variable selection, KNIFE and DOSK both perform very well, and are significantly better than the other methods.

\begin{table}[htb]
\centering{}{\small{}}%
\begin{tabular}{c|c|cccc}
\hline
{\small{}$p_0$} & {\small{}Method} & {\small{}Train MCR} & {\small{}Test MCR} & {\small{}TP} & {\small{}FN}\tabularnewline
\hline
\multirow{5}{*}{{\small{}0}} & {\small{}$L_2$ Kernel} & {\small{}2.94 (0.93)} & {\small{}2.92 (0.50)} & {\small{}-} & {\small{}-}\tabularnewline
\cline{2-6}
 & {\small{}SIS} & {\small{}2.94 (0.93)} & {\small{}2.92 (0.50)} & \textbf{\small{}1} & \textbf{\small{}0}\tabularnewline
\cline{2-6}
 & {\small{}RFE} & {\small{}2.94 (0.93)} & {\small{}2.92 (0.50)} & \textbf{\small{}1} & \textbf{\small{}0}\tabularnewline
\cline{2-6}
 & {\small{}KNIFE} & {\small{}4.00 (2.92)} & {\small{}4.32 (3.94)} & {\small{}0.98} & \textbf{\small{}0}\tabularnewline
\cline{2-6}
 & {\small{}DOSK} & {\small{}1.63 (0.73)} & \textbf{\small{}1.72 (0.34)} & \textbf{\small{}1} & \textbf{\small{}0}\tabularnewline
\hline
\multirow{5}{*}{{\small{}4}} & {\small{}$L_2$ Kernel} & {\small{}1.63 (0.89) } & {\small{}6.68 (0.75)} & {\small{}-} & {\small{}-}\tabularnewline
\cline{2-6}
 & {\small{}SIS} & {\small{}2.31 (1.22)} & {\small{}5.23 (1.50)} & \textbf{\small{}1} & {\small{}0.69}\tabularnewline
\cline{2-6}
 & {\small{}RFE} & {\small{}9.48 (12.84)} & {\small{}12.02 (12.40)} & {\small{}0.8} & {\small{}0.36}\tabularnewline
\cline{2-6}
 & {\small{}KNIFE} & {\small{}3.33 (1.30)} & {\small{}3.31 (0.50)} & \textbf{\small{}1} & \textbf{\small{}0}\tabularnewline
\cline{2-6}
 & {\small{}DOSK} & {\small{}2.07 (0.12)} & \textbf{\small{}2.02 (0.56)} & \textbf{\small{}1} & \textbf{\small{}0}\tabularnewline
\hline
\multirow{5}{*}{{\small{}8}} & {\small{}$L_2$ Kernel} & {\small{}0.08 (0.21)} & {\small{}15.07 (1.89)} & {\small{}-} & {\small{}-}\tabularnewline
\cline{2-6}
 & {\small{}SIS} & {\small{}0.96 (1.00)} & {\small{}9.53 (4.45)} & \textbf{\small{}1} & {\small{}0.66}\tabularnewline
\cline{2-6}
 & {\small{}RFE} & {\small{}5.42 (8.97)} & {\small{}12.18 (9.16)} & {\small{}0.86} & {\small{}0.46}\tabularnewline
\cline{2-6}
 & {\small{}KNIFE} & {\small{}3.48 (1.87)} & {\small{}3.89 (2.97)} & {\small{}0.99} & \textbf{\small{}0}\tabularnewline
\cline{2-6}
 & {\small{}DOSK} & {\small{}1.58 (1.63)} & \textbf{\small{}1.79 (0.34)} & \textbf{\small{}1} & \textbf{\small{}0}\tabularnewline
\hline
\end{tabular}
\caption{Results of Classification Example 1. The numbers in parentheses show
the corresponding standard deviations. MSC stands for Mis-Classification Rate, TP and FN represent true positive rates and false negative rates, respectively.}
\label{simutab3}
\end{table}

\noindent {\bf Classification Example 2:} We consider a similar example as in Classification Example 1. In particular, we let the classification signal depend on 4 predictors. For the first class, the distribution of $x_{\cdot 1}$ to $x_{\cdot 4}$ is $N\{(0,0,0,0)^T,I_4\}$. The corresponding distribution of the second class is proportional to $N\{(0,0,0,0)^T, I_4\} \mid 9< \sum_{j=1}^4 x_{\cdot j}^2 <16$. We let $p_0=0,4,8$ in this example. The classification results are reported in Table~\ref{simutab4}, and one can draw a similar conclusion as that of Classification Example 1.

\begin{table}[htb]
\centering{}{\small{}}%
{\small{}}%
\begin{tabular}{c|c|cccc}
\hline
{\small{}$p_0$} & {\small{}Method} & {\small{}Train MCR} & {\small{}Test MCR} & {\small{}TP} & {\small{}FN}\tabularnewline
\hline
\multirow{5}{*}{{\small{}0}} & {\small{}$L_2$ Kernel} & {\small{}6.34 (0.15)} & {\small{}8.08 (0.80)} & {\small{}-} & {\small{}-}\tabularnewline
\cline{2-6}
 & {\small{}SIS} & {\small{}6.34 (0.15)} & {\small{}8.08 (0.80)} & {\small{}1} & {\small{}0}\tabularnewline
\cline{2-6}
 & {\small{}RFE} & {\small{}6.34 (0.15)} & {\small{}8.08 (0.80)} & {\small{}1} & {\small{}0}\tabularnewline
\cline{2-6}
 & {\small{}KNIFE} & {\small{}7.30 (1.70)} & {\small{}8.85 (0.87)} & {\small{}1} & {\small{}0}\tabularnewline
\cline{2-6}
 & {\small{}DOSK} & {\small{}4.37 (1.74)} & \textbf{\small{}5.81 (0.73)} & \textbf{\small{}1} & \textbf{\small{}0}\tabularnewline
\hline
\multirow{5}{*}{{\small{}4}} & {\small{}$L_2$ Kernel} & {\small{}1.58 (1.08)} & {\small{}14.56 (1.23)} & {\small{}-} & {\small{}-}\tabularnewline
\cline{2-6}
 & {\small{}SIS} & {\small{}2.59 (1.02)} & {\small{}13.49 (1.87)} & {\small{}1.00} & {\small{}0.84}\tabularnewline
\cline{2-6}
 & {\small{}RFE} & {\small{}10.82 (3.96)} & {\small{}19.96 (6.87)} & {\small{}0.76} & {\small{}0.52}\tabularnewline
\cline{2-6}
 & {\small{}KNIFE} & {\small{}7.73 (1.88)} & {\small{}9.41 (1.66)} & {\small{}1} & {\small{}0}\tabularnewline
\cline{2-6}
 & {\small{}DOSK} & {\small{}4.94 (1.68)} & \textbf{\small{}6.00 (0.84)} & \textbf{\small{}1} & \textbf{\small{}0}\tabularnewline
\hline
\multirow{5}{*}{{\small{}8}} & {\small{}$L_2$ Kernel} & {\small{}0.02 (0.01)} & {\small{}22.28 (1.65)} & {\small{}-} & {\small{}-}\tabularnewline
\cline{2-6}
 & {\small{}SIS} & {\small{}2.02 (5.64)} & {\small{}19.60 (3.72)} & {\small{}0.96} & {\small{}0.72}\tabularnewline
\cline{2-6}
 & {\small{}RFE} & {\small{}8.12 (2.10)} & {\small{}22.93 (6.21)} & {\small{}0.76} & {\small{}0.50}\tabularnewline
\cline{2-6}
 & {\small{}KNIFE} & {\small{}7.21 (1.72)} & {\small{}9.03 (1.20)} & {\small{}1} & {\small{}0}\tabularnewline
\cline{2-6}
 & {\small{}DOSK} & {\small{}5.04 (1.75)} & \textbf{\small{}5.93 (0.64)} & \textbf{\small{}1} & \textbf{\small{}0}\tabularnewline
\hline
\end{tabular}
\caption{Results of Classification Example 2. The numbers in parentheses show
the corresponding standard deviations. MSC stands for Mis-Classification Rate, TP and FN represent true positive rates and false negative rates, respectively.}
\label{simutab4}
\end{table}

Next, we would like to use simulated examples to discuss the computational complexity and the compare the runtime of DOSK with other methods. According to Algorithm 1, the linear approximation in the $\mathbf{w}$ step simplifies the original non-convex optimization problem into a quadratic programming program with linear constraints. Similar to KNIFE, the order of the computational cost per iteration of DOSK should be equivalent to that of the kernel regression using the quadratic loss. Similarly, the computational cost of DOSK would perform the same as the standard SVM using the hinge loss. In practice, the actual runtime of DOSK can depend on the number of iterations used before convergence. Therefore, a proper starting point $\mathbf{w}^{(0)}$ can save the computational time significantly.

In order to assess the actual runtime performance of DOSK, we use the same four simulated examples above and fix the noise dimension as $p_0=8$. We also include two real data applications: the CPUs and Ecoli datasets. To have a general idea of the runtime in finding the best tuning parameters, we record the average time (in seconds) that each method takes for each tuning parameter value combination. For regression examples, the linear ridge and LASSO are implemented by the R package glmnet. The $L_2$ Kernel method is also implemented by glmnet but includes some extra kernel matrix calculation. SIS, RFE and COSSO are implemented by the corresponding R packages SIS, caret, and COSSO respectively. KNIFE and DOSK are implemented using R entirely. For classification examples, $L_2$ Kernel, SIS and RFE are all primarily fitted by the R package e1071 with some extra matrix calculation. KNIFE and DOSK are implemented by a R wrapper of the Matlab package CVX to conduct the two quadratic programmings in each iteration. As to the stopping criterion, we always use the default settings when there is a corresponding R package. For KNIFE and DOSK, we set the maximum iteration number to be 300 and the stopping rule as when the $L_2$-norm of the objective function change is less than $0.001$. The average runtime of all the methods for each tuning parameter set is listed in Table \ref{tab:runtime}.

Based on the results in Table \ref{tab:runtime}, it is not surprising to see that the linear ridge and LASSO take much less time than all the other methods since the core of the package glmnet contains a set of Fortran subroutines, which is much faster than the corresponding R code. The $L_2$ kernel method, SIS, and RFE are slower not only because they have more complexity but also due to the extra matrix calculation in R. Similar arguments can also be made for these methods in classification, which are implemented by the libsvm C++ code.  The results of COSSO heavily depend on the selection of the knots number. As to KNIFE and DOSK, they perform almost equivalently in terms of computational time under both the regression and classification examples. This comparison result is consistent to our previous discussion on the comparable computational complexity.  Note that KNIFE and DOSK have long runtime under classification examples because there is some additional communication cost needed for calling the Matlab package CVX from R.

\begin{figure}[htb]
\begin{center}
\includegraphics[width=0.8\textwidth,totalheight=0.8\textwidth]{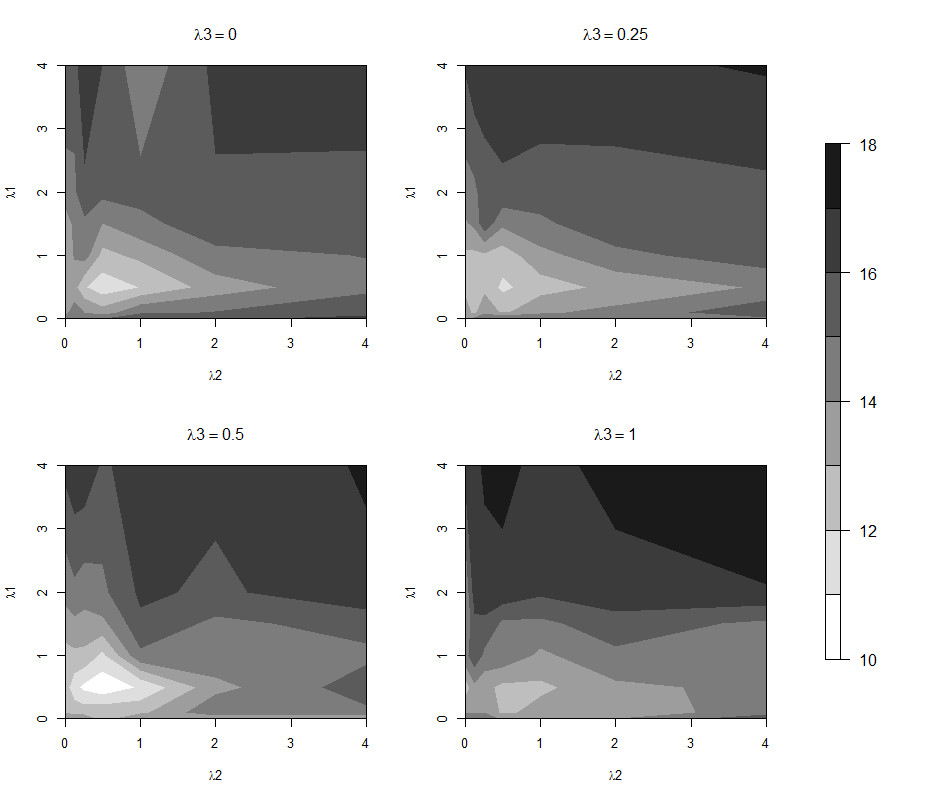}
\end{center}
\caption{Contour plots of the mean prediction errors of DOSK for Regression Example 2 where $p_0=8$. Here  $\lambda_3$ is set as $\{0, 0.25, 0.5, 1\}$ for the four panels and the kernel bandwidth $\tau=1/2\hat{\sigma}^2$, where $\hat{\sigma}$ is the median of the pairwise Euclidean distances of the simulated samples.}
\label{contour}
\end{figure}

As to the tuning parameter selection, we fix $\lambda_3=0.5$ to save the computational time. Note that there are three tuning parameters $\lambda_1, \lambda_2, \lambda_3$ in (\ref{eq:main}) for the proposed DOSK. Based on our numerical experiment, the performance of DOSK is not sensitive to the choice of $\lambda_3$, the tuning parameter for the quadratic penalty term. For illustration, we draw four contour plots of the mean prediction errors for Regression Example 2 when $p_0=8$ in Figure \ref{contour}. In particular, we set $\lambda_3$ as $\{0, 0.25, 0.5, 1\}$ respectively for each plot and calculate the optimal prediction error among all combinations of $\lambda_1$ and $\lambda_2$ with $\tau$ being $1/2\hat{\sigma}^2$, where $\hat{\sigma}$ is the median of the pairwise Euclidean distances for the simulated samples. From the result, one can observe that the best ($\lambda_1$, $\lambda_2$) combination is almost always near the coordinate $(0.5, 0.5)$ for all these $\lambda_3$ values. Because we fix $\lambda_3$ to be $0.5$ in DOSK, KNIFE and DOSK have the identical number of parameters to be tuned in practice. This choice appears to work well in all the experiments we tried. As a consequence, these two methods need approximately the same time in finding the best $\lambda$'s.

\subsection{Real Data Applications}
\label{sec:real}

In this section, we apply our DOSK method to four real data sets and explore the corresponding prediction performance. In particular, the first two real data sets are about regression problems, and the last two are for classification applications.

\noindent {\bf Regression Examples: Ozone Data and CPUs Data}

We consider the ozone pollution data in Los Angels \citep{breiman_estimating_1985}, and the Central Processing Units (CPUs) performance prediction data \citep{ein-dor_attributes_1987} as our regression applications. The ozone data set includes 330 observations, and each observation contains the daily measurement of ozone reading (the response) in 1976. Furthermore, 8 predictors that have potential impact on the ozone readings are also available, such as temperature, inversion base height, etc. The CPUs performance data set can be found in the UCI machine learning Repository \citep{Bache+Lichman:2015}. The corresponding response variable contains 209 different CPUs' published relative performance on a benchmark mix. The data set also includes 7 predictors, such as the cache size, minimum main memory, and cycle time, among others, which may be useful in predicting a computer's performance.

Before the analysis, we standardize the data sets, such that the range of each predictor is in $[0,1]$. Because we do not have separate training and testing data sets, for each replicate we randomly split the data into two equal parts, and use one for training and the other for testing. We choose the best tuning parameters in a similar way as in the simulated examples, by 5-fold cross validations on the training sets. The Laplacian kernel is used for both examples. We compare our DOSK method with LASSO, standard $L_2$ kernel learning, SIS regression with $L_2$ kernel learning, RFE with $L_2$ kernel learning, COSSO and KNIFE.

The average prediction errors in 50 replicates are summarized in Table~\ref{realtab1}. For the ozone data, the DOSK method performs better than the existing approaches in terms of the average prediction error. For the CPUs data, one can see that the standard $L_2$ kernel learning may have a potential overfitting issue, which is similar to the simulation results. In terms of variable selection, we report the predictors that are selected more than $45$ times out of the 50 replicates. In the CPUs data set, each method selects a small subset of the predictors in the models. In particular, SIS tends to fit a model with minimum main memory and maximum main memory. The RFE and LASSO approaches select maximum main memory, cache size, and maximum number of channels as the important variables. For COSSO, KNIFE and our DOSK methods, the maximum main memory and cache size are the selected variables. This is consistent with the insights given in \citet{ein-dor_attributes_1987}. In other words, to specify the performance of a computer, only a few components are necessary. Interestingly, LASSO works slightly better than SIS, RFE, or the COSSO methods in prediction. One possible explanation is that the response is not highly nonlinear in this example, and kernel learning methods without stable variable selection can lead to suboptimal results. In contrast, KNIFE performs competitively, while our DOSK enjoys the best accuracy. This suggests that variable weighted kernel learning can provide stable selection performance for real applications.


\begin{table}[t!]
\begin{centering}
\begin{tabular}{cccc|cccc}
\hline 
Examples & Reg-1 & Reg-2 & CPUs &  & Class-1 & Class-2 & Ecoli\tabularnewline
\hline 
Methods & Runtime & Runtime & Runtime & Methods & Runtime & Runtime & Runtime\tabularnewline
\hline 
Linear Ridge & 0.26 & 0.36 & 0.22 &  &  &  & \tabularnewline
LASSO & 1.12 & 0.87 & 0.57 &  &  &  & \tabularnewline
$L_{2}$ Kernel & 13.65 & 13.09 & 11.94 & $L_{2}$ Kernel & 4.39 & 4.41 & 2.18\tabularnewline
SIS & 11.18 & 13.31 & 13.50 & SIS & 17.13 & 17.91 & 13.73\tabularnewline
RFE & 41.25 & 69.27 & 57.71 & RFE & 28.42 & 39.87 & 16.72\tabularnewline
COSSO & 34.23 & 39.37 & 42.84 &  &  &  & \tabularnewline
KNIFE & 82.2 & 83.88 & 82.16 & KNIFE & 145.68 & 162.41 & 86.10\tabularnewline
DOSK & 98.46 & 97.36 & 81.25 & DOSK & 153.94 & 156.16 & 91.45\tabularnewline
\hline 
\end{tabular}
\par\end{centering}
\caption{Average runtime (in second) of each method per tuning parameter combination in the selected numerical studies.
Here $n=100$ and $p_{0}=8$ for all simulated examples.\label{tab:runtime}}
\end{table}

\begin{center}
\begin{table*}[htb]
\hfill{}
\begin{tabular}{c|cc|cc}
\hline
{\small{}} & \multicolumn{2}{c|}{{\small{}Ozone}} & \multicolumn{2}{c}{{\small{}CPUs}}\tabularnewline
\hline
{\small{}Methods} & {\small{}Train MPE} & {\small{}Test MPE} & {\small{}Train MPE} & {\small{}Test MPE}\tabularnewline
\hline
{\small{}$L_2$ Kernel} & {\small{}12.51 (1.27)} & {\small{}17.37 (1.68)} & {\small{}0.01 (0.002)} & {\small{}0.40 (0.24)}\tabularnewline
\hline
{\small{}LASSO} & {\small{}19.34 (1.36)} & {\small{}20.80 (1.69)} & {\small{}0.11 (0.04)} & {\small{}0.21 (0.09)}\tabularnewline
\hline
{\small{}SIS} & {\small{}18.72 (1.61)} & {\small{}21.47 (1.78)} & {\small{}0.11 (0.03)} & {\small{}0.33 (0.21)}\tabularnewline
\hline
{\small{}RFE} & {\small{}13.89 (1.44)} & {\small{}18.37 (1.73)} & {\small{}0.02 (0.01)} & {\small{}0.35 (0.20)}\tabularnewline
\hline
{\small{}COSSO} & {\small{}17.56 (2.14)} & {\small{}20.45 (1.96)} & {\small{}0.12 (0.07)} & {\small{}0.28 (0.12)}\tabularnewline
\hline
{\small{}KNIFE} & {\small{}11.03 (1.09)} & {\small{}17.08 (1.90)} & {\small{}0.10 (0.01)} & {\small{}0.17 (0.08)}\tabularnewline
\hline
{\small{}DOSK} & {\small{}11.21 (1.41)} & \textbf{\small{}16.92 (1.65)} & {\small{}0.09 (0.02)} & \textbf{\small{}0.16 (0.10)}\tabularnewline
\hline
\end{tabular}
\hfill{} \caption{The mean prediction error (MPE) for the ozone and CPUs data sets. }
\label{realtab1}
\end{table*}
\end{center}

\noindent {\bf Classification Examples: Breast Cancer Wisconsin Data and Ecoli Data }

For classification applications, we use the diagnostic Wisconsin breast cancer data set \citep{street_nuclear_1993}
and the Ecoli data set \citep{k_nakai_expert_1991} for illustration. These two data sets can also be found in the UCI machine learning Repository. The breast cancer data set has diagnosis results (malignant or benign) for 569 patients. The data also contain 30 predictors computed from a digitized image of a fine needle aspirate of a breast bass, such as mean distances from center to points on the perimeter, standard deviation of gray-scale values, etc. The Ecoli data set has 8 categories of proteins, and we use two categories, namely, cytoplasmic proteins and inner membrane proteins without signal sequence, for demonstration in our analysis. The total number of samples of these two classes is 220, and the data set includes 7 predictors, such as different measures of signal protein sequence recognition, consensus sequence score, amino acid content in certain outer proteins, among others.

We use DOSK with the SVM hinge loss, and compare our method with standard $L_2$ kernel SVM, SIS, RFE and KNIFE. Similar to the regression examples, we standardize all the predictors before our analysis. Furthermore, we randomly split the data sets into two equal parts, and use one for training (5 fold cross validations to select the best tuning parameters) and the other for testing. We report the average prediction error rates for various methods in Table~\ref{realtab2}, and one can see that the standard kernel SVM with the $L_2$ norm penalty can have a potential overfitting issue on these two data sets, which is consistent with the simulation results. Compared with other methods, our DOSK performs competitively.

\begin{center}
\begin{table*}[htb]
\hfill
\begin{tabular}{c|cc|cc}
\hline
{\small{}} & \multicolumn{2}{c|}{{\small{}Breast Cancer}} & \multicolumn{2}{c}{{\small{}Ecoli}}\tabularnewline
\hline
{\small{}Methods} & {\small{}Train MCR} & {\small{}Test MCR} & {\small{}Train MCR} & {\small{}Test MCR}\tabularnewline
\hline
{\small{}$L_2$ Kernel} & {\small{}0.39 (0.24)} & {\small{}7.78 (1.42)} & {\small{}0.22 (0.33)} & {\small{}13.24 (4.42)}\tabularnewline
\hline
{\small{}SIS} & {\small{}1.27 (0.73)} & {\small{}4.20 (1.09)} & {\small{}0.95 (0.68)} & {\small{}2.13 (1.21)}\tabularnewline
\hline
{\small{}RFE} & {\small{}1.33 (0.56)} & {\small{}4.26 (1.00)} & {\small{}0.95 (0.68)} & {\small{}2.13 (1.25)}\tabularnewline
\hline
{\small{}KNIFE} & {\small{}1.77 (0.54)} & {\small{}4.04 (0.78)} & {\small{}1.69 (0.81)} & {\small{}2.26 (1.27)}\tabularnewline
\hline
{\small{}DOSK} & {\small{}2.40 (0.60)} & \textbf{\small{}3.97 (1.11)} & {\small{}1.52 (1.02)} & \textbf{\small{}1.95 (1.02)}\tabularnewline
\hline
\end{tabular}
\hfill{} \caption{The Mis-Classification Rate (MCR, in percentages) for the breast cancer and Ecoli data sets. }
\label{realtab2}
\end{table*}
\end{center}

\section{Discussion}
\label{sec:discussion}

In this paper, we propose a new DOSK method in kernel learning that can perform variable selection and data extraction simultaneously. We show that under certain conditions, the new DOSK method can achieve selection consistency, and the estimated function can converge to the underlying function with a fast rate. We also develop an efficient algorithm to solve the corresponding optimization, which is guaranteed to converge to a local optimum. Numerical results show that our DOSK method is highly competitive among existing approaches.

As a remark, our DOSK method can be generalized to alleviate the computational burden for applications with massive data sets. Without loss of generality, take regression as an example. Suppose one needs to estimate a nonlinear underlying function, and the data set contains many observations and predictors. To perform kernel regression with such big data can be computationally inefficient. One way to circumvent this difficulty is to split the predictors into several parts or dividing the observations into several subsets, learn on each part individually, and then combine the results. In particular, each time one can perform our DOSK method on one piece of the data set. Because our DOSK method can have double sparsity in predictors and dual variables, for each sub-regression, it is possible to find a sparsely represented function that only involves a subset of observations and predictors. Then we can combine the selected observations and predictors to train for a global estimator. One can see that this approach can greatly reduce the computational time for problems with massive data sets. Further research can be pursued in this direction.

\section*{Acknowledgments}

The authors would like to thank the editor Professor George Michailidis, the associate editor, and referees for their helpful comments and suggestions. The authors were supported in part by NIH/NCI grant R01 CA-149569,  P01 CA142538, and NSF grant DMS-1407241.

\section*{Appendix}

\noindent {\bf Proof of Theorem~\ref{thm:algorithm}}. Because the objective function $\phi$ is lower bounded by zero, to prove convergence, it suffices to prove that for each step of updating, the objective function value is non-increasing. To this end, we will show that $\phi$ is non-increasing for Steps 2-4 in Algorithm 2. First, notice that for fixed $\mathbf{w}$, the corresponding objective functions in the $\balpha$ step and the $b$ step are convex. Hence, $\phi$ is non-increasing for Steps 2 and 3. We will focus on Step 4 next.

Without loss of generality, suppose that $\nabla_{\mathbf{w}} \phi( \balpha^{(t)}, b^{(t)}, \mathbf{w}^{(t-1)} ) \ne \bzero$ (otherwise, the algorithm has already converged). We will prove that the directional derivative along $\Delta \mathbf{w}$ is negative, with which one can verify that after Step 4, the objective function value would decrease. To this end, observe that Step 4(a) can be regarded as to minimize $\psi (\mathbf{w}) = h\{ g(\mathbf{w}) \}$, where $h(\cdot)$ is a convex and continuously differentiable function and $g(\cdot)$ is a convex or concave and continuously differentiable function of $\mathbf{w}$. Since both $h$ and $g$ are continuously differentiable, they are locally Lipshcitz continuous, and so is $\psi$. Furthermore, because $h$ and $g$ are convex or concave, there exists an open neighborhood of $\mathbf{w}^{(t-1)}$, $\mN(\mathbf{w}^{(t-1)})$, in which $h$ and $g$ are monotonic \citep{bertsekas_convex_2003}. Therefore, in $\mN(\mathbf{w}^{(t-1)})$, $\psi(\cdot)$ is monotonic.

Next, we prove that along the direction defined by $\Delta \mathbf{w}$, $\psi(\cdot)$ is monotonically deceasing in $\mN(\mathbf{w}^{(t-1)})$. To this end, first notice that Step 4 computes a descent direction of $\tilde{\psi}_{ \mathbf{w}^{ (t-1 )} } (\mathbf{w}) = h\{ g(\mathbf{w}^{ (t-1 )}) + \nabla g(\mathbf{w}^{ (t-1 )})^T (\mathbf{w} - \mathbf{w}^{ (t-1 )} ) \}$. Because the objective function of $\mathbf{w}^{(QP)}$ is quadratic, thus strictly convex, $\tilde{\psi}_{ \mathbf{w}^{ (t-1 )} } (\mathbf{w})$ is strictly decreasing along $\Delta \mathbf{w}$ within $\mN(\mathbf{w}^{(t-1)})$. Next, by similar arguments as in the proof of Proposition 1 in \citet{allen_automatic_2012}, one can verify that $\psi(\cdot)$ is monotonically deceasing along $\Delta \mathbf{w}$ within $\mN(\mathbf{w}^{(t-1)})$, and this completes the proof. \hfill $\blacksquare$

\noindent {\bf Proof of Theorem~\ref{thm:f_rate}}: Before we present our proof, we first give some lemmas.

\begin{lemma}
Suppose Assumptions 1-7 are valid. With $\lambda_1$, $\lambda_2$ and $\lambda_3$ as in Theorem~\ref{thm:f_rate}, we have that $\|\hat{\balpha}\|_1 = O_P\{ \log(n) \}$ and $|\hat{b}| = O_P\{ \log(n) \}$.
\label{lemma:01}
\end{lemma}

\noindent {\bf Proof of Lemma~\ref{lemma:01}}: With $\balpha=\bzero$ and $b = 0$, we have $ \phi( \bzero ,0,\mathbf{w}) = \frac{1}{n} \sum_{i=1}^n L(y_i,0) \rightarrow E\{L(Y,0)\}$ as $n \rightarrow \infty$, which is a constant. On the other hand, $\hat{\balpha}$ and $\hat{b}$ are (part of) the solution to the objective function in (\ref{eq:main}). Hence,
\begin{align*}
\lambda_{1} \|\hat{\balpha}\|_1 & \le \frac{1}{n} \sum_{i=1}^{n}L\big\{y_{i}, \sum_{j=1}^{n}K_{\hat{\mathbf{w}}}(x_{i},x_{j})\hat{\alpha}_{j}+\hat{b}\big\}
+\lambda_{1}\| \hat{\balpha} \|_{1}+\lambda_{2}\| \hat{\mathbf{w}} \|_{1}
+\lambda_{3}\hat{\balpha}^{T}K_{\hat{\mathbf{w}}}\hat{\balpha} \\
& \le \phi( \bzero ,0,\mathbf{w}).
\end{align*}
Consequently, we have $\|\hat{\balpha}\|_1 = O_P\{ \log(n) \}$. For $|\hat{b}|$, in regression, because the fitted function $\hat{f}$ cannot be uniformly larger or smaller than the observed responses, we have that $|\hat{b}|$ is at most $O_P(\|\hat{\balpha}\|_1)$, which is $O_P\{ \log(n) \}$ (notice that we have assumed that the error term in regression are bounded for now). For classification problems, similar arguments hold ($\hat{f}$ cannot be uniformly positive or negative, otherwise the classification problem is of less interest), and $|\hat{b}| = O_P\{ \log(n) \}$. This completes the proof. \hfill $\square$

\begin{lemma}
Suppose Assumptions 1-7 are valid. We have that $\| f_{\balpha_n^*, b_n^*} - f_0 \|_{2} = O_P  \{  \log(n)/n \}$.
\label{lemma:02}
\end{lemma}

\noindent {\bf Proof of Lemma~\ref{lemma:02}}: Notice that $\gamma_j$'s are constants, and the kernel function $K_{\mathbf{w}^*}$ is Lipshcitz by Assumption 2. Hence, we have
\begin{align*}
  & | f_{\balpha_n^*, b_n^*} (\cdot) - f_0 (\cdot) |\\
= & | \sum_{j=1}^m \gamma_j \{K_{\mathbf{w}^*} (\bx_j,\cdot) - K_{\mathbf{w}^*} (\bz_j,\cdot)\} | \\
= & O_P(\max_j \|\bx_j-\bz_j\|_2),
\end{align*}
and the goal is to prove that $\|\bx_j-\bz_j\|_2 = O_P  \{  \log(n)/n \}$ for all $j$. To this end, note that $\pr(\|\bx_j-\bz_j\|_2 > d) = (1-P_d)^n$, where $d$ is a small positive number, and $P_d = \pr(\|\bz-\bz_j\|_2 \le d) = \int_{\|\bz-\bz_j\|_2 \le d} dP$. Using Assumption 1, one can verify that we can choose $d =  2\log(n) / n $, such that $\pr(\|\bx_j-\bz_j\|_2 > d) = O_P(n^{-2})$. By the Borel--Cantelli Lemma, we have $\|\bx_j-\bz_j\|_2 = O_P  \{  \log(n)/n \}$ holds. This completes the proof. \hfill $\square$

The next lemma generalizes a theoretical result from the margin-based classifier literature to broader ranges of learning problems. In particular, in \citet{MLUM}, it was shown that the convergence rate of excess risks for margin-based classifiers is related to the convergence rate of the estimated learning function. In Lemma~\ref{lemma:03}, we extend the discussion to more general situations, in which one uses differentiable loss functions to measure the goodness of fit of $\hat{f}$.

\begin{lemma}
Suppose Assumptions 1-7 are valid. Moreover, consider a loss function $\ell\{u(f,y)\}$ that is second order differentiable with respect to $u$, where $u(f,y)$ is a function of the response $y$ and the learning function $f$. Assume that $u$ has second order derivative with respect to $f$, and the two second order derivatives are both bounded. Then we have that, if the function $f^*$ minimizes $E(\ell)$,
\begin{align*}
| E[\ell\{u(Y,f)\}] - E[\ell\{u(Y,f^*)\}] | = O\{(\|f-f^*\|_{2})^2\},
\end{align*}
and if $f^*$ is not the minimizer of $E(\ell)$,
\begin{align*}
| E[\ell\{u(Y,f)\}] - E[\ell\{u(Y,f^*)\}] | = O\{(\|f-f^*\|_{2})\}.
\end{align*}
\label{lemma:03}
\end{lemma}

\noindent {\bf Proof of Lemma~\ref{lemma:03}}: This proof is analogous to that of Theorems 5 and 6 in \citet{MLUM}. Hence, for brevity, we only list the key steps. The first step is to introduce the idea of Bregman divergence. In particular, for a convex differentiable function $g(\cdot)$, its Bregman divergence $d_g$ is defined as $d_g(f_1,f_2) = g(f_2)-g(f_1)-g'(f_1)(f_1-f_2)$. Then, one can prove that the conditional excess risk $E[\ell\{u(Y,f)\}] - E[\ell\{u(Y,f^*)\}] \mid_{\bX=\bx}$ equals to the Bregman divergence $d_{\ell}\{f^*(\bx),f(\bx)\}$. See the proof of Theorem 4 in \citet{MLUM} for more details. Combining this result with Assumption 3, we can show, in a similar manner as in the proof of Theorems 5 and 6 in \citet{MLUM}, that the claim of Lemma~\ref{lemma:03} holds. \hfill $\square$

We are ready to prove Theorem~\ref{thm:f_rate}. The proof follows a similar line as that of Theorem 1 in \citet{zhang_quantile_2015}. Therefore, we only list out the key steps here. The first step is to decompose the excess risk into two parts, the estimation error and the approximation error. In particular, let $f_{\blambda}$ be the best prediction function with respect to the penalized loss function for fixed $\blambda = (\lambda_1, \lambda_2,\lambda_3)$, i.e., $f_{\blambda} = \arginf_{f} [E \{L(Y,f) \}  + \lambda_{1}\|\mathbf{\boldsymbol{\alpha}}\|_{1}+\lambda_{2}\|\mathbf{w}\|_{1}
+\lambda_{3}\mathbf{\mathbf{\boldsymbol{\alpha}}}^{T}K_{\mathbf{w}}\mathbf{\boldsymbol{\alpha}}]$. The estimation error is defined as $E \{L(Y,\hat{f}) \} - E \{L(Y,f_{\blambda}) \}$, and the approximation error is defined to be $E \{L(Y,f_{\blambda}) \} - E \{L(Y,f_0) \}$.

Next, consider the function space $\hat{f}$ lies in, and denote it by $\mF_{\blambda}$. Define $g_f(\cdot) = s^{-1} \{L(\cdot,f) - L(\cdot, f_{\blambda})\}$, where $s$ is chosen such that the $L_2$ diameter of $\mG = \{g_f: f\in \mF_{\blambda}\}$ is $1$. Using Lemma~\ref{lemma:01}, one can verify that $s=O_P\{\log(n)\}$. From Lemma 2 in \citet{zhang_quantile_2015}, we have that the upper bound of the $L_2$ entropy number of $\mG$, $\log[N\{\eta, \mG, L_2(T_X) \}]$, is of the order $O_P(\eta^{-2})$ \citep[see, for example,][for introduction of the entropy numbers]{EPbook}. Here $T_X$ is the empirical measure of a training set, and the $L_2$ norm is $\|f\|_{L_2 (T_X)} = \{ n^{-1}\sum_{i=1}^n |f(\bx_i,y_i)|^2 \} ^{1/2}$. Consequently, one can obtain that the estimation error is of the order $O_P\{ \log(n) / \sqrt{n} \}$, by similar arguments as in the proof of Theorem 1 in \citet{zhang_quantile_2015}. Therefore, by Lemma~\ref{lemma:03}, $\|\hat{f} - f_{\blambda}\|_{2} = O_P\{ \log(n) / \sqrt{n} \}$.

On the other hand, to derive the bound for the approximation error, one can use Assumption 1, Lemmas~\ref{lemma:02} and~\ref{lemma:03}. In particular, we have that $E [ L\{ Y,  f_{\blambda} (\bX) \} ] - E [ L\{ Y,  f_0(\bX) \} ]  $ converges at a rate faster than that of $\| f_{\balpha_n^*, b_n^*} - f_0 \|_{2}^2$ (recall the definition of $f_{\blambda}$), which is $O_P  [\{  \log(n)/n \}^2 ] = O_P \{ \log^2(n)/(n^2) \}$. Thus, by Lemma~\ref{lemma:03}, we have that $\| f_{\blambda} - f_0 \|_{2} = O_P \{ \log (n)/n \}$. Consequently, one has that $\|\hat{f} - f_0\|_{2} \le  (\|\hat{f}  - f_{\blambda}\|_{2} + \| f_{\blambda} - f_0 \|_{2}) = O_P\{ \log(n) / \sqrt{n} \}$. This completes the proof. \hfill $\blacksquare$

\noindent {\bf Proof of Theorem~\ref{thm:sele_con}}: In the proof, we first assume that for regression problems, the distribution of the error has a bounded range. We will consider the more general case of sub-Gaussian distribution later.

The next lemma, Lemma~\ref{lemma:04}, is an important intermediate step to the proof of Theorem~\ref{thm:sele_con}. With Lemma~\ref{lemma:04}, we can prove that the difference between $\hat{f}$ and the best function $f_0$, in terms of the difference in their expected partial derivatives with respect to $w_j$, is converging at the rate at least $O_P\{ \log(n) / \sqrt{n} \}$. This further leads to the fact that the proposed $\lambda_2$ in Theorem~\ref{thm:sele_con} can correctly select the important variables $\bx_{(1)}$ and discard the noise $\bx_{(0)}$. Consequently, we can have the desired selection consistency for our DOSK method.

\begin{lemma}
Suppose Assumptions 1-7 are valid. With $\lambda_1$, $\lambda_2$ and $\lambda_3$ as in Theorem~\ref{thm:sele_con}, we have that for any $j=1,\ldots,p$,
\begin{align*}
\left| \left[ \frac{\partial E [L\{Y, \hat{f}(\bX)\}] }{\partial w_j} -
\frac{\partial E [L\{Y, f_0(\bX)\}] }{\partial w_j} \right] \mid_{w_j=0, \ w_i=w^*_i, \ i \ne j} \right| = O_P\left\{\frac{\log(n)}{\sqrt{n}}\right\}.
\end{align*}
\label{lemma:04}
\end{lemma}

\noindent {\bf Proof of Lemma~\ref{lemma:04}}: The proof follows a similar line as that of Theorem~\ref{thm:f_rate} and Lemma~\ref{lemma:03}. \hfill $\square$

We are ready to present the proof to Theorem~\ref{thm:sele_con}.

First, we prove that for any $j$,
\begin{align}
& \left| \left[ \frac{\partial  [\frac{1}{n} \sum_{i=1}^n L\{y_i, \hat{f}(\bx_i)\}] }{\partial w_j} -
\frac{\partial E [L\{Y, f_0(\bX)\}] }{\partial w_j} \right] \mid_{w_j=0, \ w_i=w^*_i, \ i \ne j} \right| \nonumber \\
= & O_P\left\{\frac{\log(n) \vee  \log(p) }{\sqrt{n}}\right\}.
\label{eq:a04}
\end{align}
To this end, observe that
\begin{align}
& \left| \left[ \frac{\partial  [\frac{1}{n} \sum_{i=1}^n L\{y_i, \hat{f}(\bx_i)\}] }{\partial w_j} -
\frac{\partial E [L\{Y, f_0(\bX)\}] }{\partial w_j} \right]  \right| \nonumber  \\
\le & \left| \left[ \frac{\partial  [\frac{1}{n} \sum_{i=1}^n L\{y_i, \hat{f}(\bx_i)\}] }{\partial w_j} -
\frac{\partial E [L\{Y, \hat{f}(\bX)\}] }{\partial w_j} \right]  \right| \nonumber \\
+ & \left| \left[ \frac{\partial E [L\{Y, \hat{f}(\bX)\}] }{\partial w_j} -
\frac{\partial E [L\{Y, f_0 (\bX)\}] }{\partial w_j} \right]  \right|.
\label{eq:a01}
\end{align}
As Lemma~\ref{lemma:04} bounds the second term on the RHS of (\ref{eq:a01}), we proceed to show that the first term converges at the rate $O_P [ \{ \log(n) \vee  \log(p) \} / \sqrt{n} ]$. To this end, we need to introduce the Rademacher complexity \citep{mohri2012foundations}. In particular, let $\sigma_i;~i=1,\ldots,n$ be $i.i.d.$ random variables, each taking the value $1$ with probability $1/2$, and $-1$ with probability $1/2$. Let the set of training observations $(\bx_i,y_i);~i=1,\ldots,n$, which are $i.i.d.$ from $P$, be denoted by $S$. Define the function class $\mH_n(\blambda)$ as $\mH_n(\blambda) = \{\hat{f} : \hat{f} = \argmin_{{\balpha,b,w}} \phi(\blambda)\}$, where $\phi(\blambda)$ is the objective function in (\ref{eq:main}). With $S$ fixed, we define the empirical Rademacher complexity of the function class $\mH_n(\blambda)$ as
\begin{eqnarray*}
\hat{R}_n\{\mH_n(\blambda)\} = E_{\bsigma} \{\sup_{f \in \mH_n(\blambda)} \frac{1}{n} \sum_{i=1}^n \sigma_i f(\bx_i) \},
\end{eqnarray*}
where $E_{\bsigma}$ represents the expectation with respect to $\bsigma = (\sigma_1,\ldots,\sigma_n)$. Furthermore, denote the Rademacher complexity of $\mH_n(\blambda)$ by
\begin{eqnarray*}
R_n\{\mH_n(\blambda)\} = E_S \hat{R}_n\{\mH_n(\blambda)\},
\end{eqnarray*}
where $E_S$ is the expectation with respect to the distribution of the sample $S$.

To bound the first term on the RHS of (\ref{eq:a01}), we have the following lemma.

\begin{lemma}
Suppose Assumptions 1-7 are valid. With $\lambda_1$, $\lambda_2$ and $\lambda_3$ as in Theorem~\ref{thm:sele_con}, we have that, for any $j=1,\ldots,p$, with probability at least $1-\delta$,
\begin{align}
\left| \left[ \frac{\partial  [\frac{1}{n} \sum_{i=1}^n L\{y_i, \hat{f}(\bx_i)\}] }{\partial w_j} -
\frac{\partial E [L\{Y, \hat{f}(\bX)\}] }{\partial w_j} \right]  \right|
& \le C_1 R_n\{\mH_n(\blambda)\} + T_n(\delta) \nonumber \\
& \le C_1 \hat{R}_n\{\mH_n(\blambda)\} + 3 T_n(\delta/2),
\label{eq:a02}
\end{align}
where $T_n (\delta) = C_2 \{n^{-1}\log(n) \log(1/ \delta)\}^{1/2}$, and $C_1$, $C_2$ are universal constants that are independent of $n$.
\label{lemma:05}
\end{lemma}

The proof to Lemma~\ref{lemma:05} is quite standard in the literature of Rademacher complexity. To bound the LHS of (\ref{eq:a02}) by $C_1 R_n\{\mH_n(\blambda)\} + T_n(\delta)$, one can use the McDiarmid inequality \citep{McDiarmid89} and the symmetrization technique \citep{EPbook}. To bound $C_1 R_n\{\mH_n(\blambda)\} $ by $C_1 \hat{R}_n\{\mH_n(\blambda)\} + 2 T_n(\delta/2)$, one can again use the McDiarmid inequality. See the proof of Lemma 3 in \citet{zhang_quantile_2015} for more details. Notice that there are two main differences between the proof of Lemma 3 in \citet{zhang_quantile_2015} and that of Lemma~\ref{lemma:05}. First, in \citet{zhang_quantile_2015}, the Rademacher complexity was defined on the function class $\{L(\cdot,f): f \in \mH_n(\blambda)\}$. By Talagrand's Lemma \citep[Lemma 4.2 in][]{mohri2012foundations}, the Rademacher complexity of $\{L(\cdot,f): f \in \mH_n(\blambda)\}$ can be further bounded by that of $\mH_n(\blambda)$, if the loss function $L$ is Lipshcitz. Second, the maximum change in the LHS of (\ref{eq:a02}) if we replace one $\bx_i$ or $y_i$ can be bounded by $C_3 \log(n)/n$ (this is a direct result from Lemma~\ref{lemma:01}) with $C_3$ being another constant, instead of $O(n^{-1})$ as in \citet{zhang_quantile_2015}. The rest of the proof is analogous to that of Lemma 3 in \citet{zhang_quantile_2015}, and we omit the details here. \hfill $\square$

The next step is to bound the empirical Rademacher complexity of $\mH_n(\blambda)$. To this end, recall the definition of $\tilde{f}$, and notice that
\begin{align}
E_{\bsigma} \{\sup_{f \in \mH_n(\blambda)} \frac{1}{n} \sum_{i=1}^n \sigma_i f(\bx_i) \} \le E_{\bsigma} \{\sup_{f \in \mH_n(\blambda)} \frac{1}{n} \sum_{i=1}^n \sigma_i \tilde{f}(\bx_i) \} + E_{\bsigma} \{\sup_{f \in \mH_n(\blambda)} \frac{1}{n} \sum_{i=1}^n \sigma_i b \}.
\label{eq:a03}
\end{align}
Hence, we proceed to bound the two terms on the RHS of (\ref{eq:a03}). Notice that by Lemma~\ref{lemma:01}, the first term is equivalent to $E_{\bsigma} \{\sup_{\|\tilde{f}\|_{\mH} = O_P \{\log(n)\}} \frac{1}{n} \sum_{i=1}^n \sigma_i \tilde{f}(\bx_i) \}$, and the second term is equivalent to $E_{\bsigma} (\sup_{ |b| = O_P \{\log(n)\} } \frac{1}{n} \sum_{i=1}^n \sigma_i b )$. For the first term, one can use Theorem 5.5 in \citet{mohri2012foundations} to obtain that, with Assumption 2 valid, the corresponding empirical Rademacher complexity is of the order $O_P\{ \log(n)/\sqrt{n} \}$. For the second term, notice that the distribution of Rademacher variables is similar to the binomial distribution. Therefore, we have that for large $n$, the distribution of $\sup_{ |b| = O_P \{\log(n)\} } \frac{1}{n} \sum_{i=1}^n \sigma_i b$ can be approximated by that of $|Z|$, where $ \{ C \sqrt{n}/\log(n)\} Z \sim N (0, 1)$, with $C$ a universal constant. Hence, one can verify that
\begin{align*}
E_{\bsigma} \{\sup_{ |b| = O_P \{\log(n)\} } \frac{1}{n} \sum_{i=1}^n \sigma_i b \} = E(|Z|) = O_P \{\log(n)/\sqrt{n}\}.
\end{align*}
Consequently, we have that $E_{\bsigma} \{\sup_{f \in \mH_n(\blambda)} \frac{1}{n} \sum_{i=1}^n \sigma_i f(\bx_i) \} = O_P \{\log(n)/\sqrt{n} \}$.

Next, choose $\delta = 2 p^{-1}n^{-2}$. One has that $T_n (\delta/2) = O_P [n^{-1}\log(n) \{\log(p) \vee \log(n)\} ]^{1/2}$. Consequently, with probability at least $2n^{-2}$, (\ref{eq:a03}) holds true for all the predictors. Combining this with Lemma~\ref{lemma:04} and the Borel--Cantelli Lemma, we have that (\ref{eq:a04}) is proved.

We now need to show that $\frac{1}{n} \sum_{i=1}^n L\big\{y_{i}, \sum_{j=1}^{n}K_{\mathbf{w}}(x_{i},x_{j})\alpha_{j}+b\big\}$, as a function of $( \mathbf{w}^T, \balpha^T, b )^T$, is strictly convex in a small neighborhood around $\big( (\mathbf{w}^*)^T , (\balpha_n^*)^T, b_n^* \big)^T$. Because we have shown that $f_{\balpha_n^*, b_n^*} (\bx)$ converges to $f_0$ in a rate faster than that of $\hat{f}$ to $f_0$,  this guarantees that once we arrive at a temporary point around $\big( (\mathbf{w}^*)^T , (\balpha_n^*)^T, b_n^* \big)^T$, the proposed algorithm in Section~\ref{sec:optimization} would ensure that the solution $\hat{f}$ converges to the best function $f_0$. To this end, observe that in Assumption 5, we assume that $E \left[ \frac{1}{n} \sum_{i=1}^n L \{Y_i, f (\bX_i)\} \right]$ is strictly convex. Hence, it suffices to prove that
$$\sup_{ ( \mathbf{w}^T, \balpha^T, b )^T \in \mN }  | \frac{1}{n} \sum_{i=1}^n L\big\{y_{i}, \sum_{j=1}^{n}K_{\mathbf{w}}(x_{i},x_{j})\alpha_{j}+b\big\} - E  [ \frac{1}{n} \sum_{i=1}^n L \{Y_i, f (\bX_i)\}  ]  | \rightarrow 0 $$
almost surely. Note that when $\mN$ is sufficiently small, we have $\sup_{f \in \mN} | P f | < \infty$. Moreover, by Lemma~\ref{lemma:01} and similar arguments as in the proof of Theorem 1 in \citet{zhang_quantile_2015}, one can have that the $L_2$ entropy of $\{f: f \in \mN \}$ is $\log [N\{\epsilon, \mN, L_2(P_n)\}] = O[\log\{\log(n)\}]$, where $P_n$ is the empirical measure of the training set. For any $M < \infty$, define $f_M = f \cdot I(f \le M)$, and $\mN_M = \{f_M: f \in \mN \}$. One has that $\log [N\{\epsilon, \mN_M, L_2(P_n)\}] = O[\log\{\log(n)\}]$. Therefore, by Theorem 6.2 in \citet{wellner2005empirical}, we have that $\mN$ is a $P$−-Glivenko--Cantelli class. One can then verify that this conclusion leads to that for $n$ large, $\frac{1}{n} \sum_{i=1}^n L\big\{y_{i}, \sum_{j=1}^{n}K_{\mathbf{w}}(x_{i},x_{j})\alpha_{j}+b\big\}$ is convex.

Now we have that, by Assumption 6, the partial derivative of the empirical $L$ loss with respect to each $w_j$ is such that
\begin{align*}
\frac{\partial  [\frac{1}{n} \sum_{i=1}^n L\{y_i, \hat{f}(\bx_i)\}] }{\partial w_j}  \mid_{w_j=0, \ w_i=w^*_i, \ i \ne j} \preceq O_P \left\{ \frac{\{\log(p) \vee \log(n)\}}{\sqrt{n}} \right\},
\end{align*}
for $w_j \in \mathbf{w}_{(0)}$, and
\begin{align*}
\left[ \frac{\partial  [\frac{1}{n} \sum_{i=1}^n L\{y_i, \hat{f}(\bx_i)\}] }{\partial w_j} - \frac{\partial E [L\{Y, f_0(\bX)\}] }{\partial w_j} \right] \mid_{w_j=0, \ w_i=w^*_i, \ i \ne j} \preceq O_P \left\{ \frac{\{\log(p) \vee \log(n)\}}{\sqrt{n}} \right\},
\end{align*}
for $w_j \in \mathbf{w}_{(1)}$. Because the objective function is locally convex, at the optimal point $(\hat{\mathbf{w}},\hat{\balpha}, \hat{b})$, selection consistency is equivalent to that $\lambda_2 \rightarrow 0$ at a rate no faster than $O_P \left\{ \frac{\{\log(p) \vee \log(n)\}}{\sqrt{n}} \right\}$ \citep[recall the soft thresholding rule in][]{tibshirani_regression_1996}. Hence, we have proven the selection consistency for the DOSK method under the assumption that the distribution of the error has a bounded range.

Lastly, we need to finish the proof by considering the general case that the distribution of the error in regression is sub-Gaussian. This can be done by showing that with a high probability, the actual errors would be bounded in a range. Then we can prove that the corresponding partial derivatives etc. converge at the same rate, because the probability of sub-Gaussian random variables being significantly away from $0$ converges to zero very fast, as the bound increases.

Without loss of generality, we assume that $\epsilon(\bX)$ follows a common sub-Gaussian distribution with c.d.f. $\Phi_{\epsilon}$. The generalization of this assumption to the heteroscedastic case is straightforward, because we are only concerned with the tail probability $\textrm{pr}(|\epsilon(\bX)|>t)$. Next, define $t^*=\Phi_{\epsilon}^{-1}\big( 0.5+0.5(1-\delta/2)^{1/n} \big)$, where $\delta$ is a small positive number. It can be verified that with probability at least $1-\delta/2$, all the errors $\epsilon_i;~i=1,\ldots,n$ are in $[-t^*,t^*]$.
Since $\Phi_{\epsilon}$ is the c.d.f. of a sub-Gaussian distribution with a fixed parameter, $t^*$ diverges at a rate slower than $O\{\log(n)\}$. One can check that the RHS of (\ref{eq:a01}) can be bounded similarly as in the corresponding proofs, and this completes the proof. \hfill $\blacksquare$

\noindent {\bf Proof of Theorem~\ref{thm:risk_bnd}}: The proof of this theorem is analogous to that of Lemma~\ref{lemma:05} and the second half of Theorem~\ref{thm:sele_con} (i.e., obtaining the bound on the empirical Rademacher complexity of $\mH_n(\blambda)$, as well as the convergence rate of $T_n(\delta/2)$). Therefore we omit the details here. \hfill $\blacksquare$

\newpage

\singlespace

\bibliographystyle{natbib}

\bibliography{double}

\end{document}